\documentclass[aps,floats,twocolumn,psfig,epsf]{revtex4}
\usepackage{graphicx}
\usepackage{amssymb}
\usepackage{amsmath}

\begin{document}

\title{Spectral functions of 2D systems with coupling of electrons \\
to collective or localized spin degrees of freedom}
\author{A. A. Katanin$^{a,b}$ and V. Yu. Irkhin$^{b}$}
\address{$^{a}$Max-Planck-Institut f\"ur Festk\"orperforschung, 70569, Stuttgart,Germany\\
$^{b}$Institute of Metal Physics, 620219 Ekaterinburg, Russia}

\begin{abstract}
{The spectral properties of itinerant 2D systems with (nearly) ferromagnetic
ground state are studied within the spin-fermion and the classical s-d
exchange models. While the former model describes the effect of collective
magnetic excitations on the electronic properties, the latter one considers
the effect of local moments. We use the\ equation of motion approach
combined with the $1/M$ and $1/z$ expansions ($M$ is the number of spin
components, $z$ is the coordination number) to investigate spectral
functions. In both the models the spectrum splitting occurs at low
temperatures $T$ in the renormalized classical regime due to strong magnetic
fluctuations. For the interaction $J$ between electronic and magnetic
degrees of freedom comparable to the bandwidth (the intermediate-coupling
regime) the $1/z$ expansion predicts full splitting of electronic spectrum
below some temperature $T_{s}$. At the same time, the spectrum remains only
partly split in the $1/M$ expansion with the quasiparticle\ structure which
becomes fully coherent at low }$T${.}
\end{abstract}

\maketitle

\section{Introduction}

The narrow-band d-systems pose the problem of magnetism of electronic
systems, where the electron-electron interactions are comparable to the
bandwidth. Two different points of view on magnetic excitations in such
systems can be considered. On one hand, magnetic excitations can be treated
as collective excitations of itinerant electrons\cite{Slatter}. On the other
hand, in many d- and especially f-electron systems one can consider
localized (magnetic) and itinerant excitations as two independent degrees of
freedom, interacting with each other.

The latter viewpoint is convenient, e.g., for transition metals containing
both conducting and localized electrons. For the description of such systems
the s-d exchange model was introduced about 50 years ago by Vonsovskii\cite%
{Vonsovsky}. The corresponding one-impurity model was applied later for the
explanation of the Kondo effect\cite{Kondo}; the periodic version of this
model was used for the description of the anomalous properties of
strongly-correlated f-electron systems, e.g., heavy-fermion compounds\cite%
{Hewson}. The strong-coupling (\textquotedblleft
double-exchange\textquotedblright ) limit of the s-d model was applied for
treating the interaction between localized moments and current carriers in a
narrow d-band in ferromagnetic semiconductors (e.g., chalcogenide spinels)
\cite{Nagaev} and manganates \cite{Nagaev1,Dagotto}.

The strong interaction between the itinerant and localized degrees of
freedom in the above-mentioned class of systems can lead to incoherent
structures in the spectral functions, in particular, the gap in electronic
spectrum can arise due to local correlations. Such a situation occurs e.g.
in the s-d exchange model which shows a metal-insulator transition with
increasing the interaction of conduction electrons with localized spins\cite%
{IK}. In more than three dimensions the local $d\rightarrow \infty $ picture
\cite{DMFT} is expected to give a qualitatively correct description of the
physics near the metal-insulator transition, while the situation in three
dimensions may be more involved. In particular, it was proposed that the
appreciable short-range magnetic order in $d=3$ may lead to the magnetic
spin splitting above the Curie temperature $T_{C}$ \cite{Koreman,Moriya}.
Such a splitting, which can not be described within the local theory, was
presumably observed in strong itinerant magnets Fe and Ni \cite{Mook}.

At the same time, in the peculiar $d=2$ (or quasi-two-dimensional) case
which is physically relevant for the layered magnetic systems with small
interlayer hopping (layered manganites, cuprates, and other compounds with
the perovskite structure), the strong short-range magnetic order takes
place. It is especially pronounced in the so-called \textquotedblleft
renormalized classical\textquotedblright\ temperature regime which arises
above the magnetically ordered ground state and is characterized by an
exponentially large correlation length $\xi \propto \exp (A/T)$ ($A$ is some
constant) \cite{RC}. Strong magnetic correlations in this regime result in
the structure of the electronic spectrum, which is similar to that in the
ordered phase \cite{KS,VIK,Dupuis}. As a result, a quasi-splitting of the
Fermi surfaces or a pseudogap can occur at low enough temperatures due to
ferromagnetic (FM) \cite{KKI} or antiferromagnetic (AFM)/charge density wave
(CDW) fluctuations\cite{Sadovskii}. These features of spectral functions
have clearly different nature as compared to the abovediscussed Mott gap,
since the finite ($d=2$) dimension is crucial for their appearance; they are
also expected to be relevant for quasi-2D systems with small interlayer
hopping above the magnetic transition temperature.

Recently, the pseudogap structures have been observed in layered manganite
compound La$_{1.2}$Sr$_{1.8}$Mn$_{2}$O$_{7}$\cite{LayeredM}. These
structures are present both above and below T$_{C}$ and possibly originate
from the CDW fluctuations\cite{Dagotto}. The FM fluctuations however may be
responsible for the part of the pronounced shift ($\sim $250 meV) of the
spectral weight maxima off the Fermi level above the Curie temperature.

The separation of conduction and localized magnetic degrees of freedom can
be also \textquotedblleft dynamic\textquotedblright\ and arise in purely
itinerant electronic systems due to strong short-range magnetic order. In
particular, it was pointed out that even purely itinerant electronic systems
in the regime of strong magnetic correlations can be described by an
effective classical s-d model \cite{Dupuis}.

To treat the effect of collective magnetic degrees of freedom on the
electronic properties and describe properties of 2D systems near magnetic
quantum phase transitions, another, \textit{spin-fermion model} model, was
introduced \cite{Pines,Schmalian,SF}. Although this model was originally
proposed as a phenomenological model for systems with strong
antiferromagnetic (AFM) fluctuations, the systems with strong ferromagnetic
(FM) fluctuations can be treated within this model as well \cite%
{Chubukov1,Katanin}.

The comparison of the spectral properties of systems where magnetic
fluctuations are induced by local-moments (described within the s-d model)
or collective magnetic fluctuations (described within the spin-fermion
model) in the regime of strong magnetic fluctuations is of certain interest.
The spectral properties of the 2D systems with strong magnetic fluctuations
were investigated previously within the spin-fermion model in the Eliashberg
approximation\cite{SF,Chubukov1} which neglects vertex corrections and in
the quasistatic approach, performing summation of all the types of diagrams,
but neglecting the effect of dynamic spin fluctuations \cite%
{Schmalian,Katanin}. The latter approach yields the two-peak structure of
the spectral function for both strong FM and AFM correlations with a finite
spectral weight at the Fermi level at finite $\xi $.

Contrary to the spin-fermion model, where the length of the spin vector
describing magnetic degrees of freedom at each lattice site is not fixed,
the s-d model considers fixed spin length. The result for the spectral
weight $A(\mathbf{k}_{F},\nu )\sim \nu ^{2}$ in the zero-bandwidth limit of
the spin-fermion model \cite{Schmalian,Katanin} shows that the spectral
weight vanishes at $\nu =0$ only, and a true metal-insulator transition is
not expected to occur. Such a transition is however expected to be described
by the s-d model \cite{Anokhin,AO,Zarubin}. Treating of electronic
properties near this transition is reminiscent the problem of considering
the effect of magnetic fluctuations in strongly-correlated Hubbard model.
Although a number of approaches were proposed recently which extend the $%
d\rightarrow \infty $ dynamical mean-field theory (DMFT) \cite{DMFT} to
account for the short-range \cite{DCA} and the long-range \cite%
{Kusunose,Jarrnew,DGA,Dual} correlations, the development of analytical
approaches to this problem is also of certain interest.

To study the magnetic fluctuations in the regime of the intermediate and
strong interaction between localized and itinerant subsystems, the 1/$z$
expansion, where $z$ is the coordination number, can be used. This approach
has been applied before to the ferro- and paramagnetic $s$-$d$ model (see
Refs. \cite{IK,Anokhin,AO,Zarubin}) as well as to the paramagnetic state in
the Hubbard model\cite{Anokhin,Zarubin1}. Its extension to describe the
paramagnetic state with strong magnetic correlations is of certain interest.

Another promising candidate for non-perturbative description of
strongly-correlated electronic systems is the Ward-identity approach\cite%
{HertzEdw}. Although this approach was quite successful in describing the
strongly FM ordered state \cite{HertzEdw1}, its generalization to the weak
FM and paramagnetic situation meets difficulties, since the contribution of
longitudinal spin fluctuations, which should be taken into account in these
cases, makes the system of equations for the electronic self-energy and
electron-(para)magnon vertices not closed.

This difficulty was overcome in a recently proposed combination of the
Ward-identity approach with the $1/M$-expansion \cite{KKI} ($M$ is the
number of spin components, $M=3$ for the s-d and Hubbard model). The use of
the $1/M$ expansion gives a possibility to truncate the hierarchy of
integral equations for the self-energy and electron-magnon vertices. To
apply this method to a broader class of models, we consider in the present
paper its generalization - the equation of motion approach combined with the
$1/M$ expansion. Contrary to the $1/z$-expansion, which considers equations
of motion for Hubbard operators, this method is applied to fermionic
operators, which allows us to describe both weak- and strong-coupling
regimes, as well as the crossover between them.

The plan of the paper is the following. In Sect. II we present the
theoretical models describing interaction of electronic and magnetic degrees
of freedom in strongly-correlated systems. In Sect. III we formulate the $%
1/z $ expansion for the $s$-$d$ model. In Sect. IV we use the equation of
motion method to derive the self-consistent equations for the self-energy
and electron-magnon vertices within the $1/M$ expansion, the details of the
derivation being presented in the Appendix. In Sect. V we investigate the
electronic self-energy, spectral functions and electron-magnon vertices in
different regimes. Finally, in Sect. VI we summarize the main results of the
paper.

\section{The model}

We consider a correlated electronic system with itinerant and
localized-moment subsystems, which is described by the generating functional
\begin{eqnarray}
Z[\eta ] &=&\int D[c,c^{\dagger }]D[\mathbf{S}]\exp \{-\mathcal{S}[c,\mathbf{%
S}]/T  \notag \\
&&\ -(c_{k\sigma }^{\dagger }\eta _{k\sigma }+\eta _{k\sigma }^{\dagger
}c_{k\sigma })\}  \label{Z}
\end{eqnarray}%
where the fields $c$ and $\mathbf{S}$ correspond to the electronic and spin
degrees of freedom, respectively. The action $\mathcal{S}[c,\mathbf{S}]$ has
the form
\begin{eqnarray}
\mathcal{S}[c,\mathbf{S}] &=&\text{i}T\int\limits_{0}^{1/T}d\tau \sum_{i}%
\mathbf{A}(\mathbf{S}_{i})\frac{\partial \mathbf{S}_{i}}{\partial \tau }%
+\sum_{k}(i\nu _{n}-\varepsilon _{\mathbf{k}})c_{k\sigma }^{\dagger
}c_{k\sigma }  \notag \\
&&+\sum_{q}\mathcal{R}_{q}\mathbf{S}_{q}\mathbf{S}_{-q}+J\sum_{kk^{\prime
}\sigma \sigma ^{\prime }}\mathbf{S}_{k-k^{\prime }}%
\mbox {\boldmath $\sigma
$}_{\sigma \sigma ^{\prime }}c_{k\sigma }^{\dagger }c_{k^{\prime }\sigma
^{\prime }}  \label{sd}
\end{eqnarray}%
where first three terms describe the electronic and spin subsystems, the
last term proportional to $J$ corresponds to the interaction between them; $%
q=(\mathbf{q},i\omega _{n})$, $k=(\mathbf{k},i\nu _{n})$; $\omega _{n}=2n\pi
T$ and $\nu _{n}=(2n+1)\pi T$ are bosonic and fermionic Matsubara
frequencies, $\varepsilon _{\mathbf{k}}$ is the electronic spectrum, $%
\mbox {\boldmath
$\sigma $}_{\sigma \sigma ^{\prime }}$ are the Pauli matrices, the term
proportional to $\mathcal{R}_{q}$ describes an indirect spin-spin exchange.
Below we suppose that this exchange is ferromagnetic, i.e. $\mathcal{R}_{%
\mathbf{q},0}$ has its minimum at $\mathbf{q}=0$.

Two different versions of the model (\ref{sd}) are considered in the paper.
The first version, which is referred to as \textquotedblleft s-d type
models\textquotedblright\ has the measure of integration over the field $%
\mathbf{S}$
\begin{equation}
D[\mathbf{S}]=D_{\text{sd}}[\mathbf{S}]\equiv \prod\limits_{i}\delta (%
\mathbf{S}_{i}^{2}-S^{2})d^{3}\mathbf{S}_{i},
\end{equation}%
so that the length of the field $\mathbf{S}$ is fixed to $S$. This is, in
particular, the case for the lattice analog of the standard s-d exchange
model \cite{Vonsovsky} (i.e. Kondo-lattice model). In this case the vector
potential $\mathbf{A}(\mathbf{S})$ describes the precession of free spin and
satisfies the equation $\mathbf{\nabla }_{S}\times \mathbf{A(S)}\cdot
\mathbf{S}=1$,$\ \mathbf{\nabla }_{S}=(\partial /\partial S_{x},\partial
/\partial S_{y},\partial /\partial S_{z})$. In the classical limit $%
S\rightarrow \infty $ this term fixes the field $\mathbf{S}$ to be static, $%
\partial \mathbf{S/}\partial \tau =0,$ cf. Ref. \cite{Auerbach}. For the s-d
type of models the term proportional to $\mathcal{R}_{q}$ in Eq. (\ref{sd})
can be considered as arising, e.g., from the indirect spin-spin RKKY
exchange.

The model (\ref{sd}) with $D[\mathbf{S}]=D_{\text{sd}}[\mathbf{S}]$ and $%
\mathbf{A}(\mathbf{S})=0$ (which is referred to as an effective s-d model in
the following) can be naturally obtained from the Hubbard model in the
renormalized classical regime \cite{Dupuis}. The quantity $\mathcal{R}_q$ is
determined in this mapping by the RPA (random-phase approximation)
expression for the inverse susceptibility in the ordered phase \cite{Dupuis}.

Another version of the model (\ref{sd}) we consider is the spin-fermion model%
\cite{Pines,Schmalian,SF} with
\begin{equation}
D[\mathbf{S}]=\prod\limits_{i}d^{3}\mathbf{S}_{i},\text{ }\mathbf{A}(%
\mathbf{S})=0
\end{equation}%
Contrary to the s-d model, the field $\mathbf{S}$ describes in this case
three \textit{independent } Bose-like fields, i.e. corresponding operators $%
\widehat{\mathbf{S}}_{q}^{a}$ with different $a$ commute with each other.
The spin-fermion model can be also considered as that representing the
low-energy electron and spin excitations of the Hubbard model in the
quantum-critical regime\cite{SF,IKI}. The quantity $\mathcal{R}_{q}$
corresponds in this case to the inverse susceptibility $\chi _{q}^{-1}$ and
enters as a phenomenological input parameter of the spin-fermion model.

Both the static case $\partial \mathbf{S/}\partial \tau =0$ and the case $%
\mathbf{A}(\mathbf{S})=0$ allow the generalization of the model (\ref{sd})
to $M$-component spins $\mathbf{S}_{i}=(S_{i}^{1}...S_{i}^{M})$. This
generalization will be used throughout the paper to perform the $1/M$%
-expansion of the self-energy and electron-magnon vertices.

For purely static $\mathcal{R}_{q}=\delta _{n0}\mathcal{R}_{\mathbf{q}}$ the
s-d and spin-fermion models are described by the Hamiltonian
\begin{eqnarray}
H &=&\sum_{\mathbf{k}}\varepsilon _{\mathbf{k}}\widehat{c}_{\mathbf{k}\sigma
}^{\dagger }\widehat{c}_{\mathbf{k}\sigma }+\sum_{\mathbf{q}}\mathcal{R}_{%
\mathbf{q}}\widehat{\mathbf{S}}_{\mathbf{q}}\widehat{\mathbf{S}}_{-\mathbf{q}%
}+H_{\text{int}},  \notag \\
H_{\text{int}} &=&-J\sum_{\mathbf{kk}^{\prime }}\widehat{\mathbf{S}}_{%
\mathbf{k}-\mathbf{k}^{\prime }}\sigma _{\sigma \sigma ^{\prime }}\widehat{c}%
_{\mathbf{k}\sigma }^{\dagger }\widehat{c}_{\mathbf{k}^{\prime }\sigma
^{\prime }}.  \label{HH}
\end{eqnarray}%
Here the Fourier transformed operators $\widehat{\mathbf{S}}_{i}$ which
correspond to the fields $\mathbf{S}_{i}$ in the continuum integral
formalism, obey the standard $SU(2)$ commutation relations and the condition
$\widehat{\mathbf{S}}^{2}=S(S+1)$ for the s-d model, while the components of
$\widehat{\mathbf{S}}$ commute with each other and $\widehat{\mathbf{S}}^{2}$
is not fixed for the spin-fermion model. In the present paper we discuss
only the classical limit of the s-d model with commuting $\widehat{\mathbf{S}%
}$-operators, so that the difference between these two models is only in the
additional restriction for the $\widehat{\mathbf{S}}^{2}$ for the s-d model
(the quantum s-d model with the non-commuting spin operators will be
considered elsewhere). For definiteness we assume $J>0$; the results in the
classical limit are the same for positive and negative $J$.

\section{The $1/z$ expansion for the $\mathbf{s}$-$\mathbf{d}$ model}

The $1/z$ expansion considers perturbation theory around the atomic limit of
the model ($\varepsilon _{\mathbf{k}}=\mathrm{const}$). The corresponding
sequence of equations of motion for the electronic Green's function was
considered in Refs. \cite{Anokhin,AO}. To zeroth order in $1/z$ we obtain
the result of the Hubbard-I approximation%
\begin{equation}
G_{\mathbf{k}}(\nu )=\frac{1}{F_{0}(\nu )-\varepsilon _{\mathbf{k}}}
\end{equation}%
where $F_{0}(\nu )=\nu -I^{2}/\nu$, $I=JS$. To first order in $1/z$ one has to
replace $F_{0}(\nu )\rightarrow F_{\mathbf{k}}(\nu )$ where%
\begin{equation}
F_{\mathbf{k}}(\nu )=\nu \left[ 1+\frac{J^{2}MT}{\nu }\sum\limits_{\mathbf{q%
}}\frac{\chi _{\mathbf{q}}}{F_{0}(\nu )-\varepsilon _{\mathbf{k+q}}}\right]
^{-1}  \label{Fk}
\end{equation}%
and $\chi _{\mathbf{q}}$ is the non-uniform spin susceptibility of the model
(\ref{HH}). Despite the result (\ref{Fk}) is obtained within the
strong-coupling expansion of the s-d model, it reproduces correctly in the
second order in $J$ the corresponding perturbation theory result for the
electronic self-energy \cite{IK}.

The result (\ref{Fk}) does not guarantee, however, correct analytical
properties of the electronic Green's function at arbitrary $J$. To obtain
analytical results, we consider self-consistent approximation of Hubbard-III
type \cite{Anokhin,AO}%
\begin{equation}
F_{\mathbf{k}}(\nu )=\nu \frac{1+[F_{0}(\nu )-F_{\mathbf{k}}(\nu )]%
\widetilde{G}_{\mathbf{k}}(\nu )}{1+[\nu -F_{\mathbf{k}}(\nu )]\widetilde{G}%
_{\mathbf{k}}(\nu )}  \label{Fkk}
\end{equation}%
where
\begin{equation}
\widetilde{G}_{\mathbf{k}}(\nu )=\frac{MT}{S^{2}}\sum\limits_{\mathbf{q}}%
\frac{\chi _{\mathbf{q}}}{F_{\mathbf{k}}(\nu )-\varepsilon _{\mathbf{k+q}}}.
\label{Gk}
\end{equation}%
The solution of the Eq. (\ref{Fkk})\ yields%
\begin{equation}
F_{\mathbf{k}}(\nu )=\nu +\frac{1-\sqrt{1+4I^2\widetilde{G}^2_{\mathbf{k}}(\nu
)}}{2\,\widetilde{G}_{\mathbf{k}}(\nu )}.  \label{FSE}
\end{equation}%
The result for the electronic self-energy $\Sigma _{\mathbf{k}}(\nu )=\nu
-F_{\mathbf{k}}(\nu ),$ obtained from the Eq. (\ref{FSE}) with $\chi _{%
\mathbf{q}}=S^{2}/T,$ coincides with the result of the solution of
single-impurity model of the dynamical mean-field theory (DMFT) for the
classical s-d model, the Eq. (\ref{Gk}) serving then as a self-consistency
condition.

\section{The self-consistent equations for the self-energy and vertices
within the $1/M$ expansion}

To find the electronic self-energy in the model (\ref{HH}) within the $1/M$
expansion, we consider the equation of motion for the fermionic operator $%
\widehat{c}_{\mathbf{k}\sigma }(\tau )$ in the Heisenberg representation
(see, e.g. Ref. \cite{Zagoskin})
\begin{eqnarray}
(\partial /\partial \tau +\varepsilon _{\mathbf{k}})\widehat{c}_{\mathbf{k}%
\sigma }(\tau ) &=&[H_{\text{int}},\widehat{c}_{\mathbf{k}\sigma }](\tau )
\label{Em} \\
&=&J\sum_{\mathbf{k}^{\prime }\sigma ^{\prime }}\mbox {\boldmath $\sigma $}%
_{\sigma \sigma ^{\prime }}\widehat{\mathbf{S}}_{\mathbf{k}-\mathbf{k}%
^{\prime }}(\tau )\widehat{c}_{\mathbf{k}^{\prime }\sigma ^{\prime }}(\tau ).
\notag
\end{eqnarray}%
Applying this equation to the electronic Green's function, we find the
expression for the electronic self-energy
\begin{equation}
\Sigma _{k}=\frac{1}{J}\sum_{q}{}^{\prime }(\Gamma _{k;q}^{z}+\Gamma
_{k;q}^{\perp })G_{k+q}^{0}  \label{se}
\end{equation}%
where we have used the notation
\begin{equation*}
\sum_{q}{}^{\prime }\equiv TJ^{2}\sum_{\mathbf{q}}\sum_{i\omega _{n}}.
\end{equation*}%
The vertices $\Gamma $ describe scattering of an itinerant electron by the
magnetic excitation (cf. Ref. \cite{HertzEdw}),
\begin{eqnarray}
\Gamma _{k;q}^{m} &=&\int d\tau _{1}d\tau _{2}e^{i\nu _{n}(\tau _{1}-\tau
_{2})+i\omega _{n}\tau _{2}}  \label{Gam} \\
&&\ \ \times \langle T[\widehat{c}_{\mathbf{k}\sigma }^{\dagger }(\tau _{1})%
\widehat{c}_{\mathbf{k}+\mathbf{q}\sigma }(\tau _{2})\widehat{S}_{{-\mathbf{q%
}}}^{m}(0)]\rangle G_{k}^{-1}(G_{k+q}^{0})^{-1},  \notag
\end{eqnarray}%
$T[...]$ stands for the imaginary-time ordering, $G_{k}^{0}$ and $G_{k}$ are
the bare and full fermionic Green's functions, respectively, $m=+,-,$ or $z$%
. Note that, despite the similarity of the definition of the vertex (\ref%
{Gam}) with that used for the Hubbard model in Ref. \cite{HertzEdw}, $%
\widehat{S}_{\mathbf{q}}^{m}$ denotes in our case the spin-subsystem
operator rather than the spin-density operator of itinerant electrons.

The vertices (\ref{Gam}) can be found by applying equation of motion (\ref%
{Em}) once more. For the following consideration, it is convenient to
introduce the corresponding one-particle irreducible vertex with amputated
bosonic and fermionic leg (cf. Ref. \cite{HertzEdw})
\begin{equation}
\gamma _{k;q}^m=G_{k+q}^0\Gamma _{k;q}^m/(J\chi _q^mG_{k+q})  \label{ga}
\end{equation}
where $\chi _q^m$ is the dynamic spin susceptibility. The equation of motion
relates the vertices (\ref{ga}) to the one-particle irreducible vertices of
the interaction of the electron with two (para-) magnons. This vertex is in
turn related to the vertex of an electron interaction with three paramagnons
and so on (see Appendix for additional details). In the following we denote
the vertex of an interaction of an electron with $n$ paramagnons as $\gamma
_{k;q_1...q_n}^{m_1...m_n}$ where $m_i=+,-,z$ are the paramagnon spin
indices, $k$ is the vector of the incoming fermion momentum and frequency, $%
q_1...q_n$ are the vectors of the paramagnon momenta and frequencies. To
truncate the resulting hierarchy of equations for the vertices, we follow
the approach of Ref. \cite{KKI} and retain only the terms which contribute
to the self-energy in the zeroth and first order in $1/M$ (see Appendix).

The resulting system of equations for the self-energy and vertices reads
\begin{eqnarray}
\Sigma _{k} &=&M\sum_{q}{}^{\prime }\gamma _{k+q}G_{k+q}\chi _{q}
\label{Eqs} \\
\gamma _{k} &=&1+\sum_{q}{}^{\prime }\left[ (2-M)\gamma
_{k+q}^{2}G_{k+q}^{2}+\gamma _{k+q}^{zz}G_{k+q}\right] \chi _{q}  \notag \\
\gamma _{k}^{zz} &=&M\sum_{q}{}^{\prime }\left[ 2\gamma
_{k+q}^{3}G_{k+q}^{3}+\gamma _{k+q}\gamma _{k+q}^{zz}G_{k+q}^{2}\right.
\notag \\
&&\left. +\gamma _{k+q}^{zz\perp }G_{k+q}\right] \chi _{q}+\alpha \gamma
_{k}G_{k}  \notag \\
\gamma _{k}^{zz\perp } &=&-2M\sum_{q}{}^{\prime }\left[ \gamma
_{k+q}^{4}G_{k+q}^{4}+\gamma _{k+q}^{2}\gamma _{k+q}^{zz}G_{k+q}^{3}\right.
\notag \\
&&\left. +\gamma _{k+q}\gamma _{k+q}^{zz\perp }G_{k+q}^{2}\right] \chi
_{q}-\alpha \gamma _{k}^{2}G_{k}^{2},  \notag
\end{eqnarray}%
where $G_{k}=[i\nu _{n}-\varepsilon _{\mathbf{k}}-\Sigma _{k}]^{-1}$ is the
full (dressed) electronic Green's function, $\gamma
_{k}^{m_{1}...m_{n}}=\gamma _{k;0...0}^{m_{1}...m_{n}}$ are the
electron-paramagnon vertices at zero paramagnon momenta, $\gamma _{k}=\gamma
_{k;0}^{z}=\gamma _{k;0}^{\bot }$,\ $\chi _{q}=\chi _{q}^{z},$ $\alpha =0$
for the spin-fermion model and $\alpha =-2$ for the s-d model.

Comparing Eqs. (\ref{Eqs}) with the result (\ref{FSE}) of the $1/z$
expansion we see that these expansions treat vertex corrections in different
ways. While the $1/M$ expansion treats accurately the momentum dependence of
the vertices, the "average" vertex of the $1/z$ expansion $[\sqrt{1+4I^2%
\widetilde{G}^2_{\mathbf{k}}(\nu )}-1]/[2I^2 \widetilde{G}^2_{\mathbf{k}%
}(\nu )]$ interpolates between the weak- and strong coupling regimes.
Both approaches, however, are expected to be applicable in the weak- and
intermediate coupling regime. Below we consider the result of application of
these approaches to calculation of spectral functions in the s-d and
spin-fermion models.

\section{The results for the spectral functions, self-energy, and vertices}

Equations (\ref{Eqs}) give a possibility to investigate the evolution of the
spectral properties with varying electron-spin coupling $J$ and the strength
of magnetic correlations. For practical calculations, one has to specify an
explicit form of the magnetic susceptibility $\chi _{q}$. In the following
we consider the low-temperature regime with strong magnetic fluctuations,
where we employ an ansatz
\begin{equation}
\chi _{{q}}=\frac{1}{M}\frac{A}{\mathbf{q}^{2}+\xi ^{-2}}\delta _{n0},\text{%
~~}|\mathbf{q}|\ll 1  \label{hi11}
\end{equation}%
for the classical s-d exchange model and
\begin{equation}
\chi _{{q}}=\frac{1}{M}\frac{A}{\mathbf{q}^{2}+\xi ^{-2}+r|\omega _{n}|/|%
\mathbf{{q}|}},\text{~~}|\omega _{n}|/v_{F}\ll |\mathbf{q}|\ll 1
\label{hi12}
\end{equation}%
for the effective s-d and the spin-fermion models (we have picked out
explicitly the factor $1/M$ for further convenience). While the constant $A$
is determined by the stiffness of spin excitations and can be arbitrary, the
correlation length $\xi $ for the s-d type models is chosen to fulfill the
sum rule%
\begin{equation}
MT\sum\limits_{q}\chi _{{q}}=S^{2} \label{sum_rule}
\end{equation}%
For the spin-fermion model the correlation length is an independent
parameter, which, however, can be related to the quantity $\Delta $
according to%
\begin{equation}
MT\sum\limits_{\mathbf{q}}\chi _{(\mathbf{q},0)}=(\Delta /J)^{2}\label{sum_rule1}
\end{equation}%
For the ferromagnetically ordered ground state $\Delta $ is expected to be
almost temperature independent at low $T$.
We also consider the
high-temperature regime with $\chi _{{q}}=S^{2}/(MT)\delta _{n0}$.

Note that the ansatz (\ref{hi11}) and (\ref{hi12}) neglects the
non-analytical corrections to the spin susceptibility \cite{Nonan} which are
expected to be not too important in the renormalized classical regime.

\subsection{$T=0$ and high-temperature results}

We start investigating the solutions of Eqs. (\ref{Gk}) and (\ref{Eqs}) from
the $\xi \rightarrow \infty $ ($T\rightarrow 0$)\ limit. In this case the
dominant contribution to momenta sums comes from the vicinity of $q=0$
point. Neglecting the momentum and frequency transfer $q$ in the Green's
functions and the vertices in the Eqs. (\ref{Gk}) and (\ref{Eqs}) we obtain
for the s-d model in the both $1/z$ and $1/M$ expansion,
\begin{equation}
\Sigma _{k}=\frac{I^{2}}{\overline{\nu }},\,\,\gamma _{k}=1-\frac{%
I^{2}}{\overline{\nu }^{2}}  \label{NS}
\end{equation}%
where $\overline{\nu }=\nu -\varepsilon _{\mathbf{k}}+i0$.
The result (\ref{NS}) with $\overline{\nu }=\nu $ coincides with that in the
atomic limit of the classical s-d model, where the spin dynamics can be
indeed neglected and there are two energy levels at $\nu =\pm I$
for parallel and antiparallel projections of the spins of itinerant electron
and localized moment. In the $\xi \rightarrow \infty $ case the two poles of
the Green's function correspond to the splitting of the Fermi surface by
strong magnetic fluctuations.

The result (\ref{NS}) can be compared to the corresponding result of the $%
1/M $ expansion for the spin-fermion model \cite{KKI}:
\begin{eqnarray}
\Sigma _{k} &=&\frac{M\,(\Delta _{0}^{2}+\overline{\nu }^{2}-\sqrt{\overline{%
\nu }^{2}-\alpha _{1}\Delta _{0}^{2}}\,\sqrt{\overline{\nu }^{2}-\alpha
_{2}\Delta _{0}^{2}}\,)}{2\,\left( 2+M\right) \,\overline{\nu }}  \notag \\
\gamma _{k} &=&\frac{M}{2\,\left( 2+M\right) ^{2}\Delta _{0}^{2}\overline{%
\nu }^{2}}\,[2\overline{\nu }^{4}+(6\,+M)\Delta _{0}^{2}\overline{\nu }^{2}
\notag \\
&&-M\Delta _{0}^{4}\ +(M\Delta _{0}^{2}-2\overline{\nu }^{2})\sqrt{\overline{%
\nu }^{2}-\alpha _{1}\Delta _{0}^{2}}  \notag \\
&&\,\times \sqrt{\overline{\nu }^{2}-\alpha _{2}\Delta _{0}^{2}}],
\label{SF}
\end{eqnarray}%
where $\Delta _{0}=\Delta(T\rightarrow 0),$ $\alpha _{1,2}=1+4(1\pm \sqrt{1+M/2})/M$; the branch Im$\sqrt{z}\geq 0$
of the square roots is chosen to guarantee the correct analytical properties
for $\Sigma $ and $\gamma .$ The result (\ref{SF}) yields also the two-peak
structure of the spectral function with peaks at $\nu \simeq \pm \Delta _0$ and a finite gap $\Delta _{sf}\sim
\Delta _{0}/2$ at the Fermi level. As discussed in Refs. \cite{KKI,Katanin},
the finite gap in the spectral function in this case is however an artifact
of the first-order approximation in $1/M$, since the actual spectral
function in the atomic limit has a behavior at small frequencies $A(\nu )\sim |\nu |^{M-1}$ 
which is non-analytic in $1/M$ (see Refs. \cite{Schmalian,Katanin}). However, as it will be shown
below, the gap disappears quickly at finite $\xi $, making the situation in
this case more favorable for the application of $1/M$ expansion.

Now we study the solutions to the equations (\ref{Eqs}) at finite $\xi $
(and $T$). At high temperatures the correlation length is small and the
q-dependence of the magnetic susceptibility can be neglected, $\chi _{{q}%
}=S^{2}/(MT)\delta _{n0}.$ The criterion for this behavior is $T\gg
J^{2}/v_{F}$ for the s-d exchange model and $T\gg v_{F}$ for the
spin-fermion model ($v_{F}$ is the Fermi velocity). The latter criterion is
never fulfilled in real systems (which is a consequence of the fact that the
local moments are not formed in the spin-fermion model, i.e. the magnetism
formation mechanism is essentially Stoner-like), so that we consider the s-d
model only. In the abovementioned limit we obtain in the $1/z$ expansion
\begin{subequations}
\label{DMFT}
\begin{eqnarray}
\Sigma _{\nu } &=&[\sqrt{1+4 I^2 (G_{\nu }^{\text{loc}})^{2}}%
-1]/(2G_{\nu }^{\text{loc}})  \label{DMFTa} \\
G_{\nu }^{\text{loc}} &=&\int\nolimits_{-\infty }^{\infty }\rho _{0}(\nu
^{\prime })(\nu -\nu ^{\prime }-\Sigma _{\nu })^{-1}  \label{DMFTb}
\end{eqnarray}%
where $\rho _{0}(\nu )$ is the non-interacting density of states for the
dispersion $\varepsilon _{\mathbf{k}}$. As mentioned in Sect. III, Eq. (\ref%
{DMFTa}) coincides with the result of the solution of the corresponding
impurity model of DMFT, while Eq. (\ref{DMFTb}) is the self-consistency
condition. The result (\ref{DMFT}) is correctly reproduced by the $1/M$
expansion in the $d\rightarrow \infty $ limit which is obtained by replacing
the non-local Green's functions by the local one, $G_{\nu }^{\text{loc}%
}=\sum_{\mathbf{k}}G_{k}.$ At finite $d$ this expansion leads, however, to
different result, which we do not consider in this paper.

The evolution of the self-energy $\Sigma _{\nu }$, the spectral function at
the Fermi surface $A(\nu )=-{\rm Im}$ $G_{\mathbf{k}_{F},\nu }/\pi ,$ and
the density of states $\rho (\nu )=-{\rm Im}$ $G_{\nu }^{\text{loc}}/\pi $
with increasing $I$ calculated according to Eq. (\ref{DMFT}) for the Bethe
(semielliptic) bare density of
states $\rho _{0}(\nu )=2\sqrt{D^{2}-\nu ^{2}}/(\pi D^{2})$
is shown in Fig. 1 ($D$ is half of the bandwidth). Although the
spectral functions have one-peak structure at small enough $I$,
one can observe that the quasiparticle picture is in fact invalid at
arbitrarily small $I$ since the real part of the self-energy has
positive slope and $|{\rm Im}\Sigma _{\nu }|$ is maximum at the Fermi
level. These features are very similar to those observed earlier in the
spin-fermion model in the presence of strong magnetic correlations \cite%
{Katanin}, although their physical origin in the present case is different.
Above a critical value $I>I _{c}=0.5D$ both the real and
imaginary parts of the self-energy diverge at the Fermi level and the
spectral functions have the two-peak structure. The density of states $\rho
(\nu )$ acquires a gap at the Fermi level above $I _{c}$, which
corresponds to a metal-insulator transition.
\begin{figure}[tbp]
\includegraphics[width=8cm]{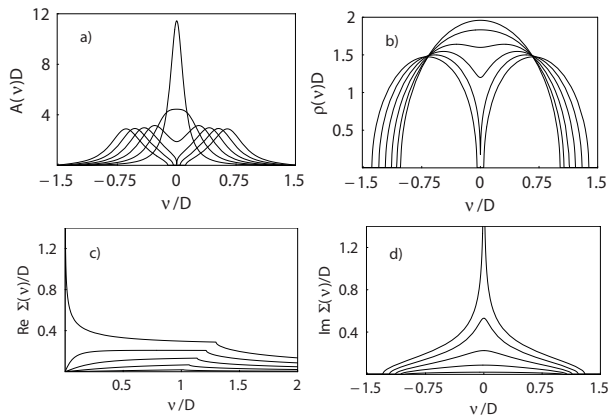} 
\caption{The spectral functions (a), the interacting density of states $%
\protect\rho (\protect\nu )$ (b), and real and imaginary parts of the
self-energy (c,d) of the s-d model in the high-temperature regime at $I=0.1nD
$, $n=1...5$ (a,b) (from below to above), $n=2...7$ (c) and $n=1...6$ (d)
(from above to below). }
\end{figure}

\subsection{Low-temperature results in the weak-coupling regime}

Now we investigate the low-temperature regime with large correlation length $%
\xi $ and the susceptibility ansatz (\ref{hi11}) or (\ref{hi12}). First we consider the
contribution of the static spin fluctuations with zero bosonic Matsubara
frequencies $\omega _{n}=0$ only (so-called static approximation). The
effect of the dynamic spin fluctuations with non-zero bosonic Matsubara
frequencies, which are less important in the renormalized classical regime,
is considered below. In the static approximation and for the nearly constant
Fermi velocity $\mathbf{v}_{F}$ in the vicininity of the Fermi momentum $%
\mathbf{k}_{F}$ both the self-energy and vertices are actually functions of $%
i\nu _{n}-\varepsilon _{\mathbf{k}}$ only:
\end{subequations}
\begin{equation}
\Sigma _{k}=\Sigma (i\nu _{n}-\varepsilon _{\mathbf{k}});~\gamma
_{k}^{m_{1}..m_{n}}=\gamma ^{m_{1}..m_{n}}(i\nu _{n}-\varepsilon _{\mathbf{k}%
})  \label{lin}
\end{equation}%
The integration over momenta can be then simplified by introducing an
auxiliary variable $a=v_{F}\mathbf{q(k}_{F}\mathbf{/}k_{F})$ (cf. Ref. \cite%
{Katanin}),
\begin{eqnarray}
&&T\sum_{\mathbf{q}}(\gamma _{k+\mathbf{q}}^{m_{1}..m_{n}})^{p}G_{k+\mathbf{q%
}}^{n}\chi _{\mathbf{q}}%
\begin{array}{c}
=%
\end{array}%
\frac{AT}{4\pi M}\int\limits_{-D}^{D}\frac{da}{\sqrt{a^{2}+v_{F}^{2}\xi
^{-2}}}  \notag \\
&&\ \ \ \ \ \ \ \ \ \ \ \ \ \ \ \ \ \ \ \ \ \ \ \ \ \ \ \times \frac{\lbrack
\gamma ^{m_{1}..m_{n}}(\nu -a)]^{p}}{[\nu -a-\Sigma (\nu -a)+i0]^{n}}
\label{m1}
\end{eqnarray}%
where we have performed analytical continuation $i\nu _{n}\rightarrow \nu +i0
$ and restricted the integartion over $a$ by $\pm D$ to account for the
effect of finite band width.

To solve Eqs. (\ref{Gk}), (\ref{FSE}), and (\ref{Eqs}) numerically we
parametrize the energy dependence of the self-energy $\Sigma (\nu )$ and
vertices $\gamma ^{m_{1}..m_{n}}(\nu )$ by a set of values at the points \{$%
\nu _{r}$\}$_{r=1}^{N_{e}}$ suitably chosen on the real axis (we choose $%
N_{e}\sim 800\div 1000$) and iterate these equations with some initial
condition until the convergence is reached. In the following we put $A=1/v_F$
for definiteness.

\begin{figure}[tbp]
\includegraphics[width=8cm]{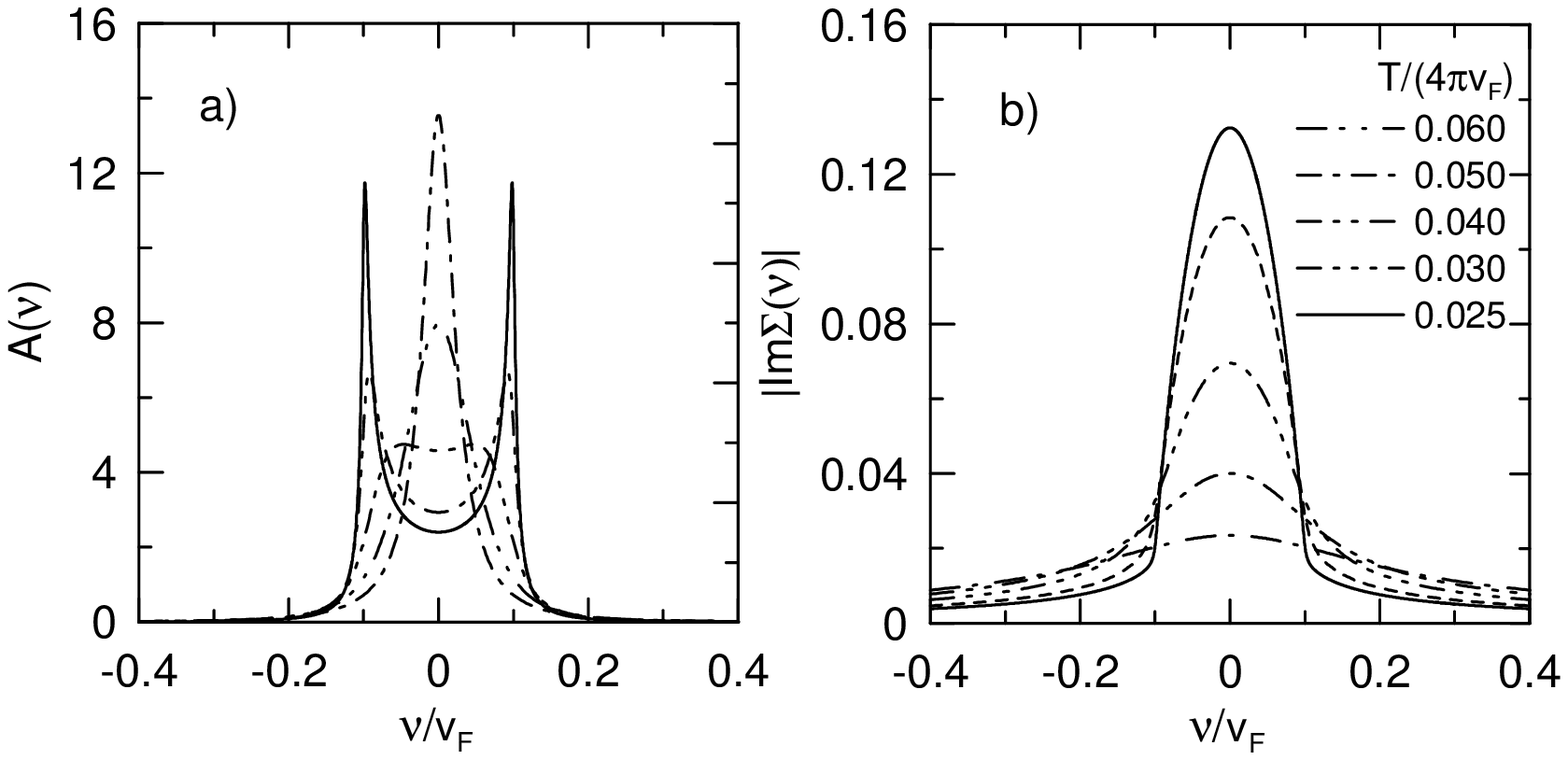} 
\caption{Spectral functions (a) and the imaginary part of the self-energy
(b) of the classical s-d model in the weak-coupling regime ($I=0.1 v_F, D=5
v_F$) at different low temperatures within the $1/z$-expansion}
\label{Fig:Fig4a}
\end{figure}

First we consider the results of the solution of these equations for the s-d
model. The frequency dependence of the self-energy $\Sigma _{\nu }$ and the
spectral functions within the $1/z$ expansion is shown in Fig. 2, the
corresponding results for the self-energy, specral functions and the vertex $%
\gamma _{\nu }$ within the $1/M$ expansion are shown in Fig. 3.

One can observe that the form of the spectral functions in $1/z$ and $1/M$
expansions is similar. In particular, we find that the imaginary part of the
self-energy has a Lorentz-like form with $|$Im$\Sigma (0)|\sim T\xi ,$ the
real part of the self-energy acquires a large positive slope at the Fermi
level, $\partial $Re$\Sigma /\partial \nu \sim T\xi ^{2},$ so that the
quasiparticle picture is invalid. With decreasing inverse correlation length
to the values $\xi _{r}^{-1}\sim \Delta _{0}/v_{F}$, the structure of the
spectral functions changes from the one-peak to two-peak form. At $\xi
^{-1}<\xi _{r}^{-1}$ the electronic spectrum and corresponding Fermi-surface
are quasi-split due to strong magnetic fluctuations.

These results are similar to the previously obtained results for the
spin-fermion model within the quasistatic approach, Ref. \cite%
{Schmalian,Sadovskii,Katanin}. The results of the $1/M$-expansion for the
spin-fermion model are shown in Fig. 4.
At large $\xi $ we obtain $\Delta
=[(ATJ^{2}/4\pi)\ln (D\xi /v_{F})]^{1/2}$; this quantity approaches
a constant value $\Delta _{0}$ (which is the analog of the interaction $I$ in the s-d model)
in the $T\rightarrow 0$ limit in the renormalized-classical regime.
The qualitative behavior of the
self-energy, spectral functions and vertices is very similar to those of
Fig. 3 for the s-d model and the results of Ref. \cite{Katanin}. In
particular, the spectral functions have two-peak form at small enough $\xi
^{-1}$ with the peaks being, however, wider than found for the s-d model.

\begin{figure}[tbp]
\includegraphics[width=8cm]{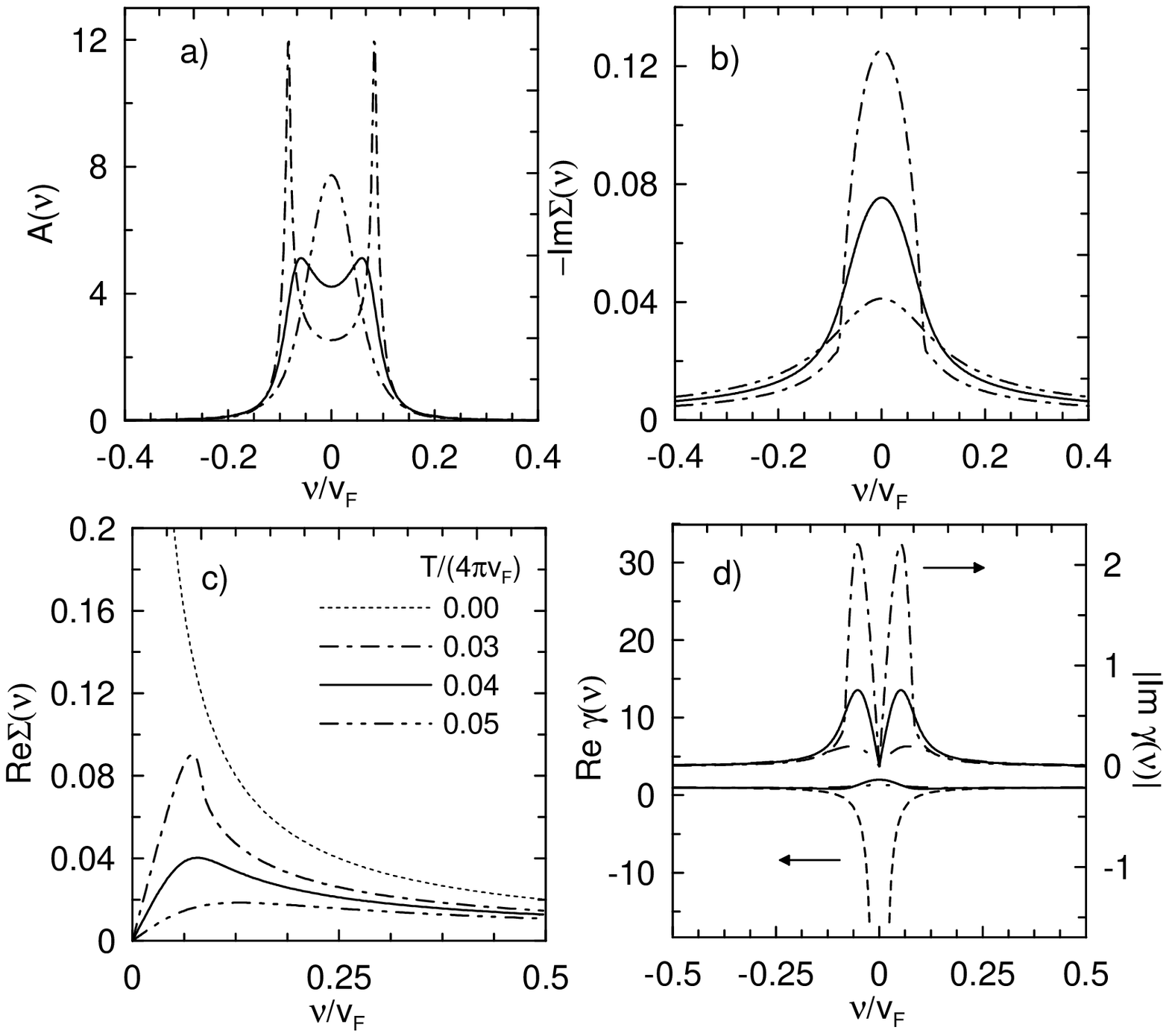} 
\caption{The spectral functions (a), the imaginary and real parts of the
self-energy (b,c), and the vertex function $\protect\gamma $ (d) of the s-d
model in the weak-coupling regime in the first order $1/M$-expansion at $M=3$%
, $I=0.1 v_F$, $D=5 v_F$ and different values of teperature.}
\label{Fig:Fig3}
\end{figure}

Now we consider the effect of the dynamic spin fluctuations with nonzero
bosonic Matsubara frequencies. In the approximation of constant Fermi
velocity $\mathbf{v}_{F}$ the self-energy and vertices depend on the difference $i\nu
_{n}-\varepsilon _{\mathbf{k}}$. After calculation of the self-energy at the
imaginary-frequency axis we use the Pad\'{e} approximants to perform
analytical continuation of the results to the real axis \cite{Pade}.
Alternatively, one can perform calculations directly on the real axis, the
results of both approaches being almost identical. The integration over
momenta in Eqs. (\ref{Eqs}) can be performed similar to Eq. (\ref{m1}),
\begin{eqnarray}
&&T\sum_{\mathbf{q}}\sum_{i\omega _{n}}(\gamma _{k+\mathbf{q}%
}^{m_{1}..m_{n}})^{p}G_{k+q}^{n}\chi _{q}%
\begin{array}{c}
=%
\end{array}%
\frac{AT}{4\pi M}  \notag \\
&&\ \times \sum_{i\omega _{n}}\int\limits_{-D}^{D}da\frac{f_{\omega
_{n}}(a)[\gamma ^{m_{1}...m_{n}}(i\nu _{n}+i\omega _{n}-a)]^{p}}{[i\nu
_{n}+i\omega _{n}-a-\Sigma (i\nu _{n}+i\omega _{n}-a)]^{n}}  \notag \\
f_{\omega _{n}}(a) &=&\frac{2}{\pi }\int\limits_{|a|/v_{F}}^{D/v_{F}}\frac{%
qdq}{\sqrt{q^{2}-(a/v_{F})^{2}}}\frac{A}{q^{2}+\xi ^{-2}+r|\omega _{n}|/q}
\label{sum1}
\end{eqnarray}

\begin{figure}[tbp]
\includegraphics[width=8cm]{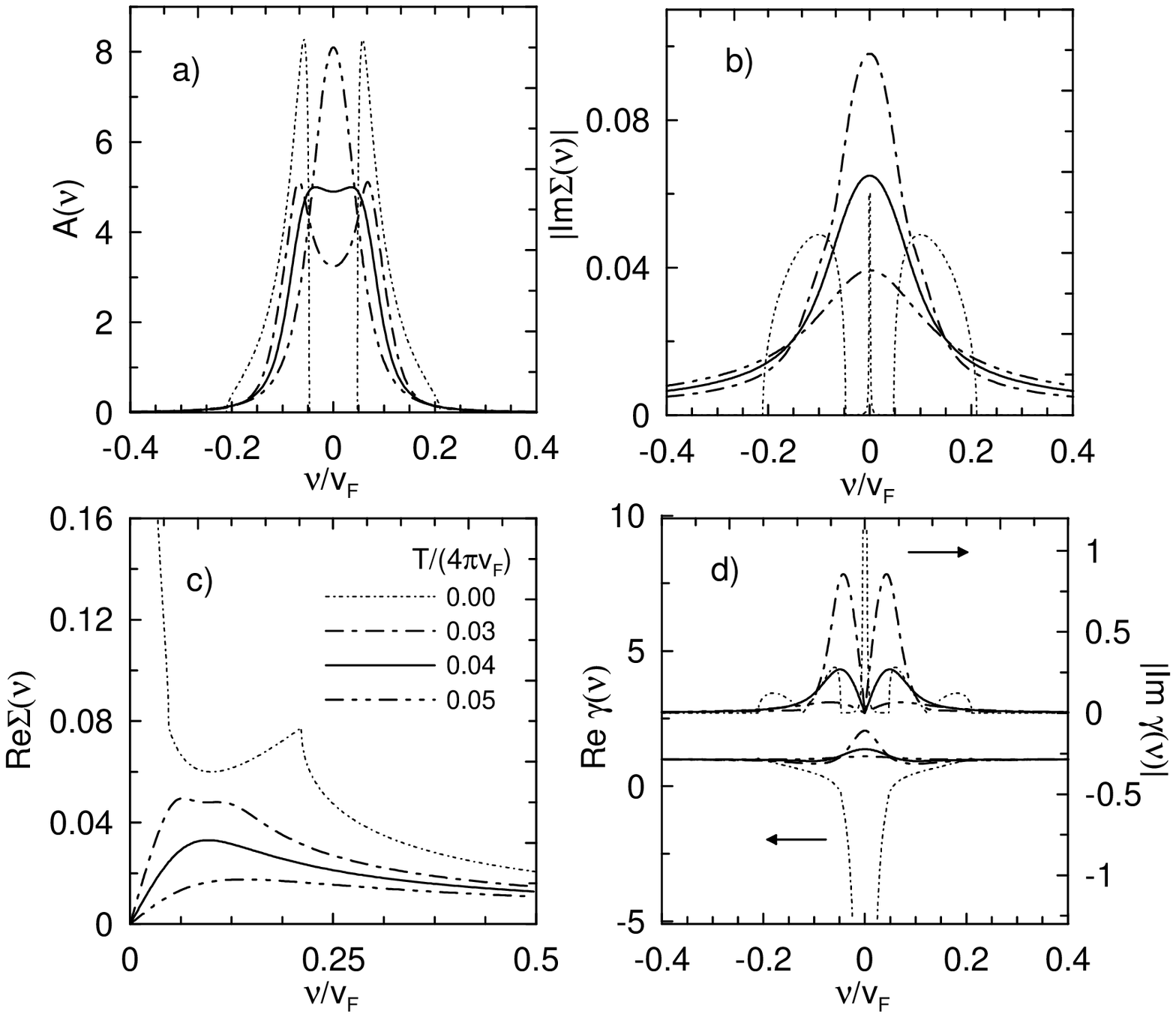} 
\caption{The same as Fig. 3 for the spin-fermion model with $J=0.2 v_F$, $\Delta_0=0.1 v_F$.}
\label{Fig:Fig4}
\end{figure}

The results for the self-energy and spectral functions in the renormalized
classical regime of the effective s-d model with different bandwidth are
shown in Fig. 5. Although the low-energy behavior of the self-energy and
spectral functions (Figs. 5a,b) is similar to that found in the static
approximation (Fig. 3), the self-energy at higher frequencies $\Delta
_{0}\ll |\nu |\ll D$ behaves as $\Sigma (\nu )\sim \nu ^{2/3}$. At very high
frequencies $|\nu |\gtrsim D$ the behavior of the self-energy $\Sigma (\nu
)\sim 1/\nu $ is restored. The results for the spin-fermion model with the
dynamic spin fluctuations are shown in Fig. 6 and have the same qualitative
frequency dependence as for the effective s-d model; the low-energy behavior
of the spectral functions (Figs. 6c,d) is close to the corresponding results
of the static approximation, Fig. 4. Therefore, the effect of dynamic spin
fluctuations in the renormalized classical regime is not pronounced, and the
static approximation describes correctly the low-energy behavior of the
self-energy and spectral functions.

A treatment of the dynamic spin fluctuations with nonzero bosonic Matsubara
frequencies is necessary to consider the self-energy and spectral functions
in the quantum critical regime. Although in that case the ansatz for the
susceptibility (\ref{hi11}) may not be correct because of the possible
non-analytic corrections \cite{Nonan}, it is still interesting to
investigate the self-energy and spectral functions in this regime with this
ansatz to clarify the role of dynamic spin fluctuations. In the
quantum-critical regime the quantity $\Delta _{0}=[(ATJ^{2}/4\pi)\ln
(D\xi /v_{F})]^{1/2}$ becomes temperature-dependent itself and one has to
employ additional ansatz for the temperature dependence of correlation
length. According to the Hertz-Millis theory \cite{Millis} (which is valid
in the absence of non-analytical corrections to susceptibility),
\begin{equation}
\xi ^{-1}=B(T/v_{F})^{1/2},
\end{equation}%
for definitness we put $B=1.$

\begin{figure}[tbp]
\includegraphics[width=8cm]{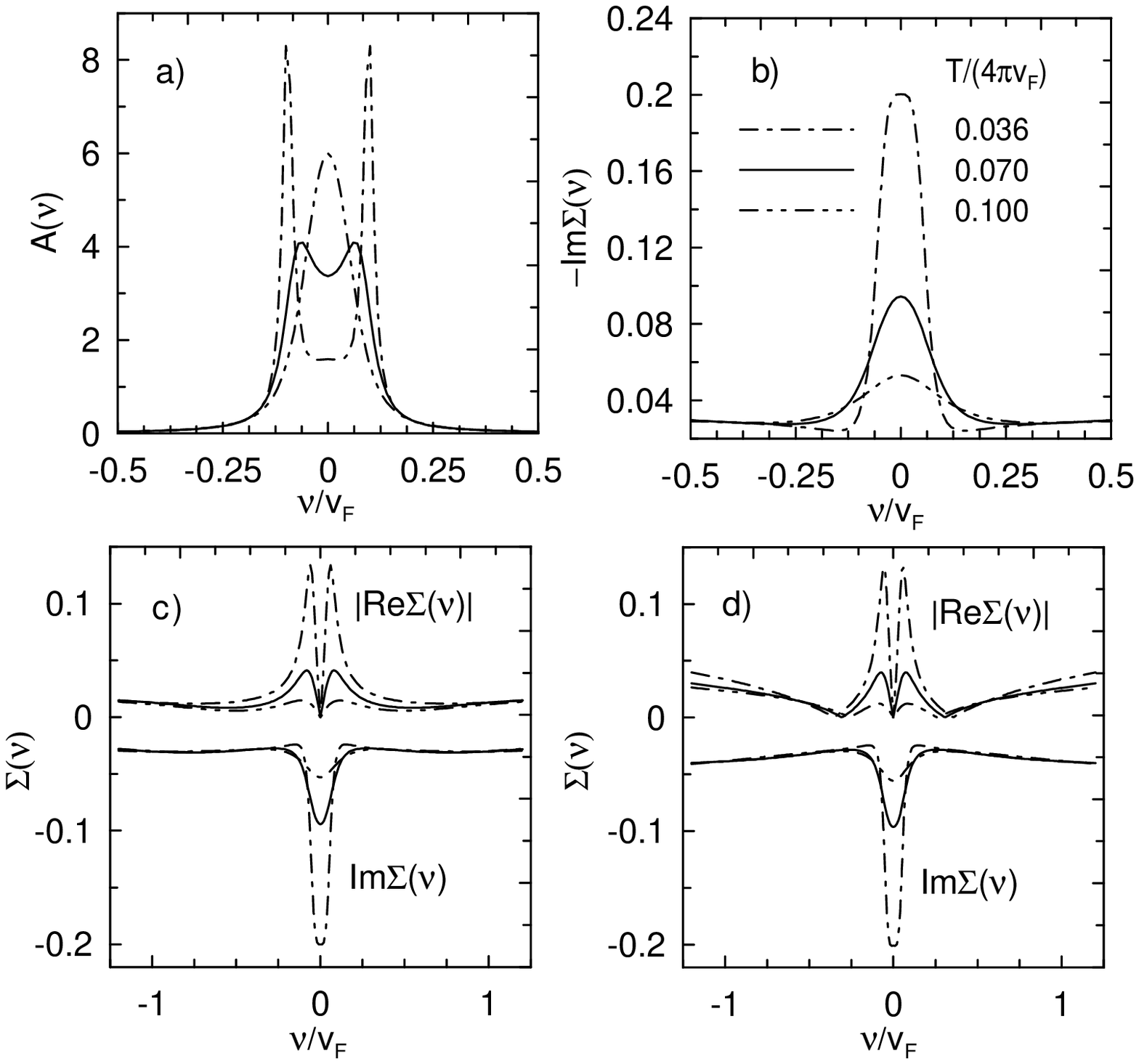} 
\caption{The spectral functions (a), the imaginary and real parts of the
self-energy (b-d) of the effective s-d model in the renormalized classical
regime with account of the dynamic spin fluctuations at the bandwidth $D=v_F$
(a-c) and $D=5 v_F$ (d), $I=0.1 v_F$, and different values of temperature.}
\label{Fig:Fig7a}
\end{figure}

\begin{figure}[tbp]
\includegraphics[width=8cm]{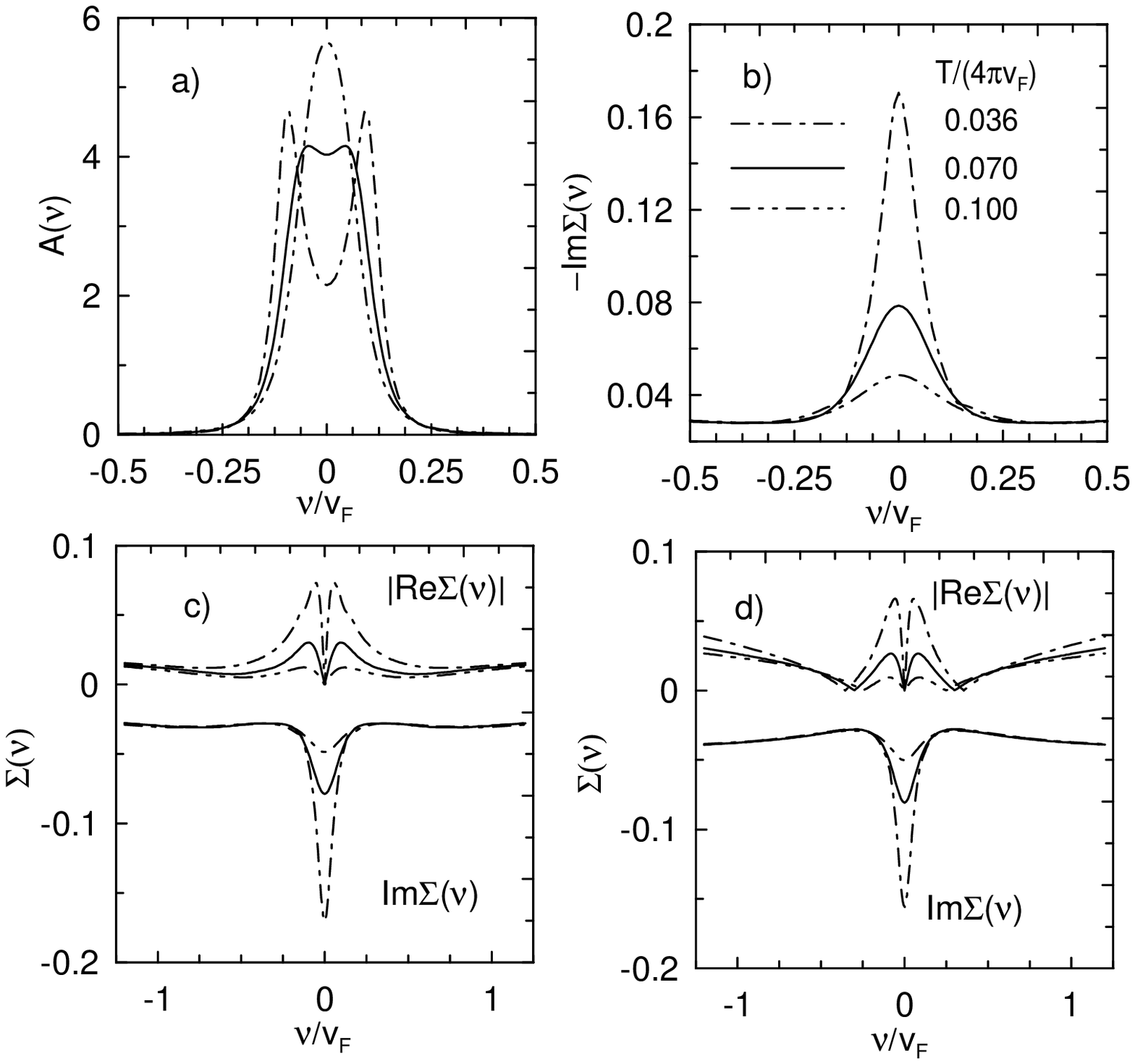} 
\caption{ The same as Fig. \protect\ref{Fig:Fig7a} for the spin-fermion
model with $J=0.2 v_F$, $\Delta_0=0.1 v_F$.}
\label{Fig:Fig8}
\end{figure}

\begin{figure}[tbp]
\includegraphics[width=8cm]{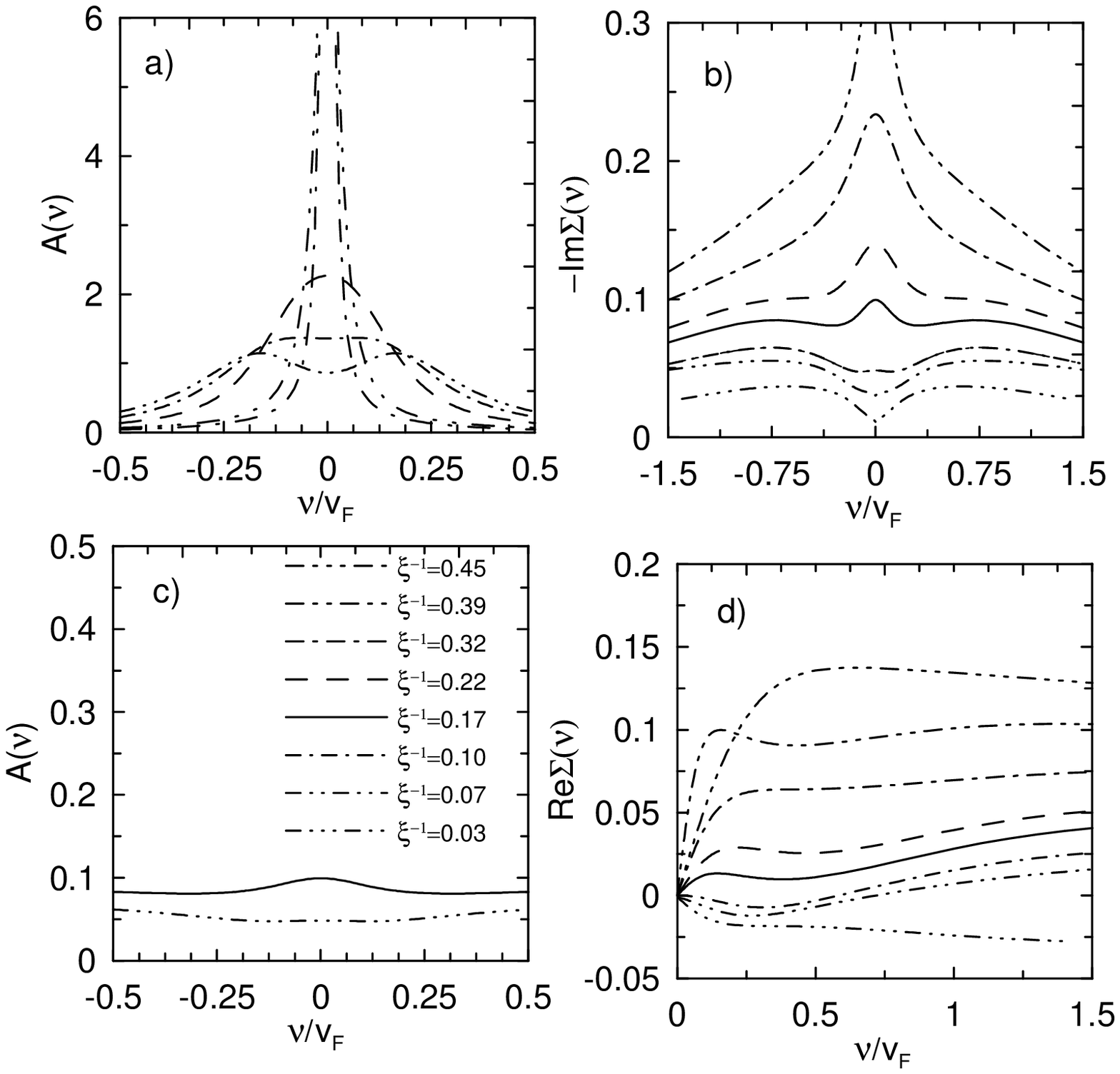} 
\caption{ The spectral functions (a,c) and the imaginary and real parts of
the self-energy (b,d) of the spin-fermion model in the quantum-critical
regime for different values of the correlation length.}
\label{Fig:Fig9}
\end{figure}

The resulting self-energy and spectral functions are shown in Fig. 7. The
frequency dependence of these quantities is determined by the interplay of
static and dynamic fluctuations. At small $\xi ^{-1}$ the dynamic
fluctuations dominate, the real part of the self-energy has negative slope
and the imaginary part has a minimum at the Fermi level, the spectral
functions having one-peak structure, so that the quasiparticle picture is
valid. However, the behavior of the self-energy $\Sigma (\nu )\sim \nu
^{2/3} $ at intermediate frequencies $(Tv_{F})^{1/2}\ll |\nu |$ $\ll D$
leads to strong damping of the quasiparticles, similar to the charge
instability case \cite{DelAnna}. At larger $\xi ^{-1}$ the static
fluctuations start to dominate and the low-frequency behavior found in Ref.
\cite{Katanin} in the quasistatic approximation is restored: the imaginary
part of the self-energy acquires maximum at the Fermi level, the real part
having a positive slope; this invalidates completely the quasiparticle
picture. Although the two-peak structure of the spectral functions is formed
at intermediate $\xi ^{-1},$ only a structure with one broad peak survives
with increasing $\xi ^{-1}$. The observed behavior of the self-energy is
also very similar to that found for the Pomeranchuk instability case in Ref.
\cite{DelAnna}. This shows that the vertex corrections (which are different
in the cases of charge and spin instabilities) are not too important in the
quantum critical regime.

\subsection{Low-temperature results in the intermediate-coupling regime}

With increasing the interaction $I$ in the s-d model one approaches the
metal-insulator transition. To obtain numerical solutions in this regime, it
appears necessary to use smooth cutoff corresponding to the finite bandwidth,
e.g. introducing the cutoff function%
\begin{equation}
w(a)=\frac{1}{2}\left( 1-\tanh \frac{2|a-D|}{D}\right)
\end{equation}%
in the integration over $a$ in the Eq. (\ref{m1}) and extending the limits
of integration from $-\infty $ to $\infty .$ The frequency dependence of the
spectral functions $A(\nu )$ and the self-energy $\Sigma _{\nu }$, obtained
from the $1/z$ expansion for $D/v_{F}=2,1,0.7$ is shown in Fig. 8. One can
see that at temperatures lower than some finite temperature $T_{s}$ (i.e. at
large enough $\xi $) the spectral functions vanish at small $\nu ,$ which
corresponds to the splitting of the electronic spectrum by magnetic
fluctuations. The finiteness of the temperature $T_{s}$ is, however, most
likely an artefact of $1/z$ expansion. The imaginary part of the self-energy
also vanihsh at small $\nu $ and $T<T_{s},$ except for the point $\nu =0,$
where it has a $\delta $-function contribution $-a\delta (\nu ),$ $a$ being
some constant.

The results of the $1/M$ expansion for the spectral functions $A(\nu )$ and
the self-energy $\Sigma _{\nu }$, and vertices $\gamma _{\nu }$, at the
Fermi surface at $M=3$ are shown in Fig. 9. One can observe that at not too
low temperatures the results of $1/M$ expansion are similar to the results
of \thinspace $1/z$ expansion. At low temperatures the $1/M$ expansion shows
finite spectral weight at the Fermi level, but has additional spikes of $%
{\rm Im}\Sigma (\nu )$ at $|\nu |=\nu _{c}\simeq \Delta _{0}$, at some
temperature $T_{s}^{\prime }$ we find ${\rm Im}\Sigma (\nu _{c})=0,\ \
{\rm Re}\Sigma (\nu _{c})=\nu _{c},$ so that the pole of the electronic
Green's function shifts to the real axis. At $T<T_{s}^{\prime }$ the
solution to the equations (\ref{Eqs}) becomes non-analytical in the upper
half-plane. Although small damping of electronic excitations is expected to
appear in higher orders of the $1/M$-expansion, the temperature $T_{s}^{\prime }$
seems to correspond to a crossover to the regime with strongly coherent
excitations at new preformed Fermi surfaces.

\begin{figure}[tbp]
\includegraphics[width=8cm]{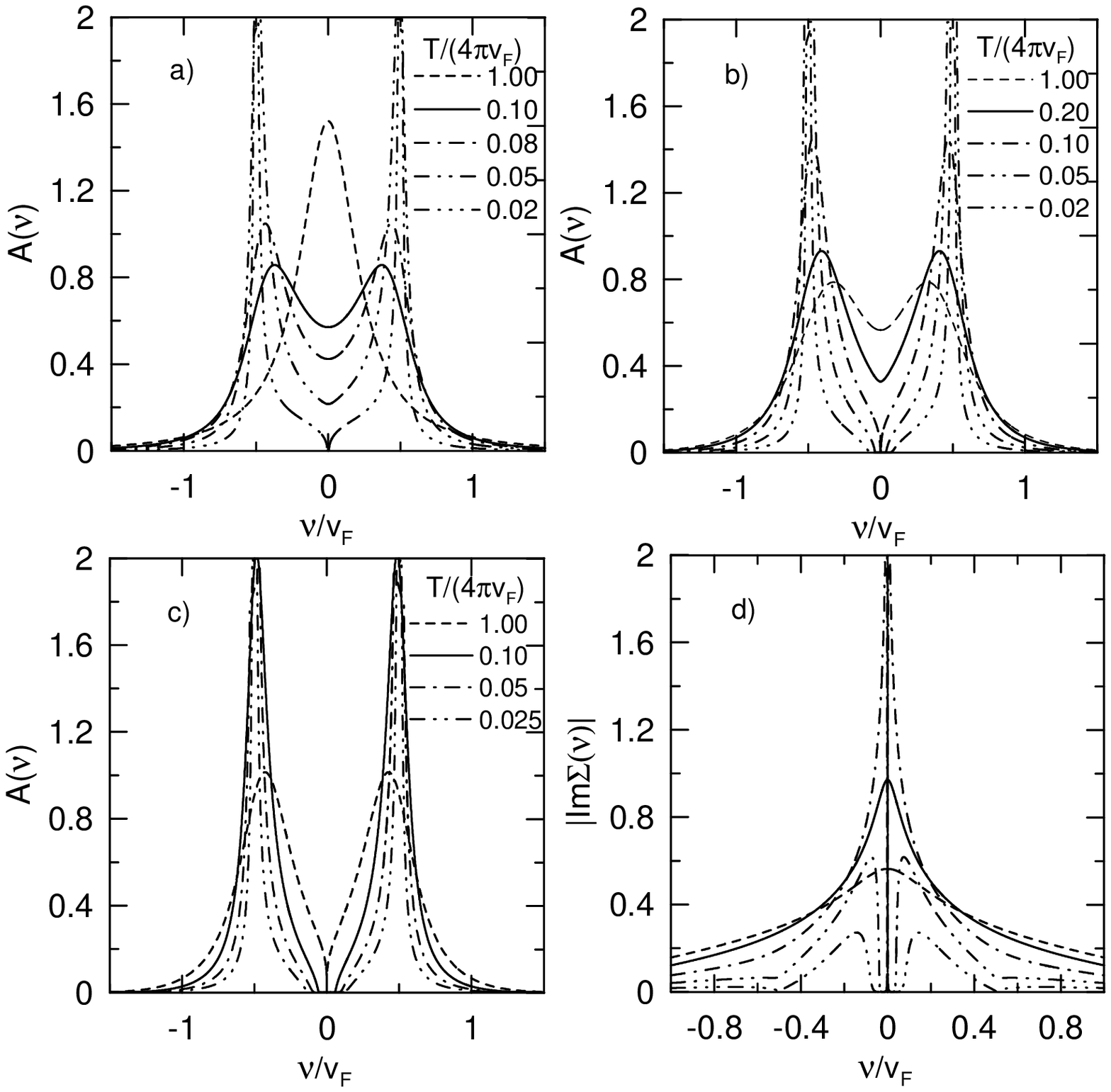} 
\caption{The spectral functions of the classical s-d model in the
intermediate-coupling regime $I=0.5 v_F$ for the bandwidth $D=2 v_F$ (a), $%
D=v_F$ (b), and $D=0.7 v_F$ (c) at different low temperatures. (d) shows the
imaginary part of the self-energy for $D=v_F$.}
\label{Fig:Fig3b}
\end{figure}

\begin{figure}[tbp]
\includegraphics[width=8cm]{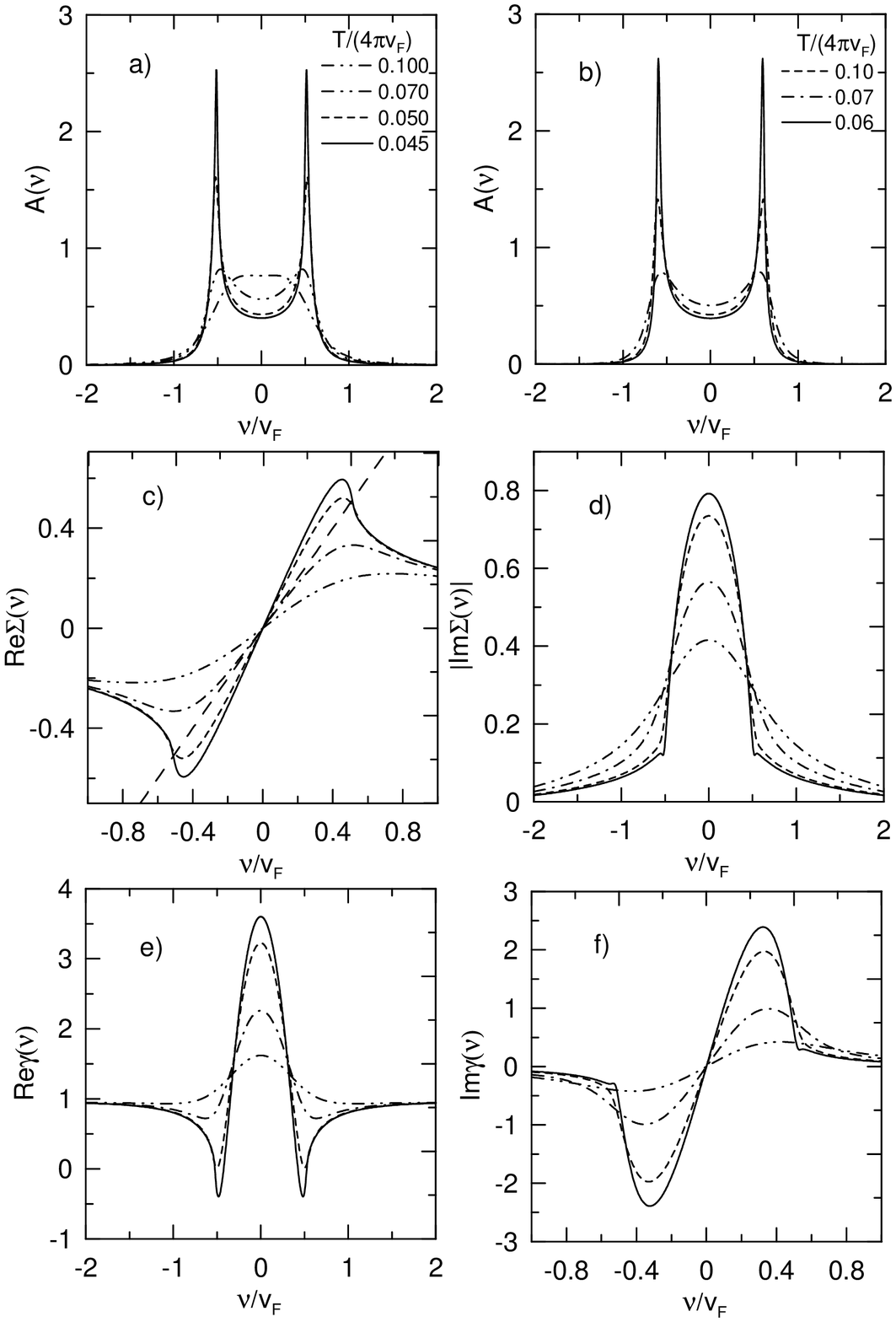} 
.
\caption{The spectral functions of the classical s-d model with the
bandwidth $D=2 v_F$ (a) and $D=v_F$ (b) in the first order $1/M$-expansion
at $M=3$ at different low temperatures and $I=0.5 v_F$. (c-f) show the
real and imaginary parts of the self-energy and the vertex function $\protect%
\gamma $ for $D=2 v_F$}
\label{Fig:Fig3a}
\end{figure}

\section{Conclusion}

We have investigated the problem of splitting of the electronic spectrum due
to coupling of electronic degrees of freedom to local magnertic moments and
collective magnetic modes in itinerant systems. In both regimes of weak- and
intermediate coupling of magnetic and electronic degrees of freedom the
spectral functions at sufficiently low temperatures have a two-peak form,
arising as a pre-splitting of the energy spectrum. In the
intermediate-coupling regime, the width of the peaks in the classical s-d
model tends to zero at the temperature $T_{s}^{\prime }$ in the first order
of the $1/M$-expansion, making the excitations at the preformed Fermi
surfaces coherent at $T<T_{s}^{\prime }$. Although a small damping of the
electronic excitations is expected to arise at finite $T$ in higher orders
of the $1/M$-expansion, it is not expected to change qualitatively the form
of the spectral functions. In $1/z$ expansion we observe the full splitting
of electronic spectra in the intermediate coupling regime at sufficiently
low temperatures.

The dynamic spin fluctuations do not change qualitatively the low-frequency
behavior of the spectral functions in the renormalized-classical regime at
weak-coupling. These fluctuations are nevertheless important in the
quantum-critical regime, where they lead to a rich behavior of the spectral
functions, which is determined by the intetrplay of static and dynamic
fluctuations.

A generalization of the results obtained in the present paper to the quantum
s-d model is of great interest. This generalization will allow to describe
analytically the formation of the Kondo resonance at the Fermi level near
the metal-insulator transition in $d \rightarrow \infty$, as well as its
interplay with magnetic degrees of freedom in finite-dimensional quantum
Kondo lattices. Another problem, which is important for the theory of
correlated magnetic metals, is the investigation of the local-moment
formation and magnetic fluctuations in itinerant-electron systems (e.g.,
described by the Hubbard model).

We are grateful to P. Igoshev for careful reading of the manuscript. The
research described was supported in part by Grant No. 4640.2006.2 (Support
of Scientific Schools) and 07-02-01264a from the Russian Basic Research
Foundation.

\section*{Appendix. The derivation of the equations for the self-energy and
vertices}

In this Appendix we consider the derivation of the system of integral
equations for the self-energy and vertices. Using the equation of motion (%
\ref{Em}) we obtain

\begin{eqnarray}
\ &&\ \ H_{0}(\partial _{x})\langle T[c_{x}c_{y}^{\dagger
}S_{a}^{m_{1}}S_{b}^{m_{2}}S_{c}^{m_{3}}...]=\delta (x-y) \\
\ \times &&\langle T[S_{a}^{m_{1}}S_{b}^{m_{2}}S_{c}^{m_{3}}...]\rangle
+J\langle T[c_{x}c_{y}^{\dagger }(\mbox {\boldmath $\sigma $}\mathbf{S}%
_{x})S_{a}^{m_{1}}S_{b}^{m_{2}}S_{c}^{m_{3}}...]\rangle  \notag
\end{eqnarray}%
where $H_{0}(\partial _{x})=\partial _{\tau }+\varepsilon (\nabla )$, $x$,$y$%
,$a$,$b$, and $c$ are the space-time coordinates. Translating this to the
momentum-frequency space and combining the result with the Dyson equation $%
G^{-1}=-(H_{0}+\Sigma )$ we find
\begin{eqnarray}
\Sigma _{k} &=&TJ\sum_{q,m}{}\Gamma _{k;q}^{m}G_{k+q}^{0};\,\,  \label{SE} \\
\Gamma _{k;q_{1}...q_{n}}^{m_{1}...m_{n};\sigma \sigma ^{\prime }}
&=&-G_{k}^{-1}\delta _{q_{1+...+q_{n}}}R_{q_{1}...q_{n}}^{m_{1}...m_{n}}-TJ
\notag \\
&&\ \times \sum_{q_{n+1},m_{n+1},\sigma ^{\prime \prime }}{}\Gamma
_{k;q_{1}...q_{n+1}}^{m_{1}...m_{n+1},\sigma \sigma ^{\prime \prime }}\sigma
_{\sigma ^{\prime \prime }\sigma ^{\prime }}^{m_{n+1}}  \notag \\
&&\ \times G_{k+q_{1}+...+q_{n+1}}^{0}  \notag
\end{eqnarray}%
where $G_{k}^{0}=(i\nu _{n}-\varepsilon _{\mathbf{k}})^{-1},$
\begin{eqnarray}
\Gamma _{k;q_{1}...q_{n}}^{m_{1}...m_{n};\sigma \sigma ^{\prime }}
&=&-G_{k}^{-1}(G_{k+q_{1}+...+q_{n}}^{0})^{-1}\int\limits_{0}^{\beta }d\tau
\cdot \,d\tau _{1}...d\tau _{n}\,  \notag \\
&&\times \langle T[S_{\mathbf{q}_{1}}^{m_{1}}(\tau _{1})...S_{\mathbf{q}%
_{n}}^{m_{n}}(\tau _{n})c_{\mathbf{k}\sigma }^{\dagger }(\tau )  \label{rmv}
\\
&&\times c_{\mathbf{k+q}_{1}+...+\mathbf{q}_{n},\sigma ^{\prime
}}(0)]\rangle e^{i\nu \tau +i\omega _{1}\tau _{1}+...+i\omega _{n}\tau _{n}}
\notag
\end{eqnarray}%
are the \textit{reducible} vertices of the interaction of an electron with $%
n $ magnons,
\begin{eqnarray}
R_{q_{1}...q_{n}}^{m_{1}...m_{n}} &=&\int\limits_{0}^{\beta }d\tau
_{1}...d\tau _{n}e^{i\omega _{1}\tau _{1}+...+i\omega _{n}\tau _{n}}  \notag
\\
&&\ \ \times \langle T[S_{\mathbf{q}_{1}}^{m_{1}}(\tau _{1})...S_{\mathbf{q}%
_{n}}^{m_{n}}(\tau _{n})]\rangle
\end{eqnarray}%
is the spin correlation function.

\begin{figure}[tbp]
\includegraphics[width=8cm]{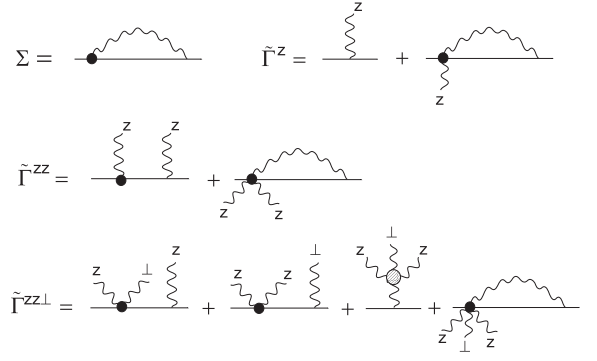} 
\caption{Diagram representation of Eqs. (\protect\ref{req}). Solid lines
correspond to the bare electronic Green functions, the black circles stand
for the one-particle reducible vertices (\protect\ref{conn}), wiggly lines
denote external legs of these vertices. The dashed circle correspond to the
vertex $\widetilde{R}_{q_1q_2q_3q_4}^{m_1m_2m_3m_4}$.}
\label{Fig:Fig10}
\end{figure}

To obtain the equations for 1-particle irreducible vertices we first define
the connected vertices via the recursive relations
\begin{eqnarray}
\widetilde{\Gamma }_{k;q_{1}...q_{n}}^{m_{1}...m_{n};\sigma \sigma ^{\prime
}} &=&\Gamma _{k;q_{1}...q_{n}}^{m_{1}...m_{n};\sigma \sigma ^{\prime }}
\label{conn} \\
&&+(G_{k}^{0})^{-1}\delta _{q_{1}+...+q_{n}}R_{q_{1}...q_{n}}^{m_{1}...m_{n}}
\notag \\
&&-\sum_{s=2}^{n-1}\sum_{\{I_{s}^{n}\}}\delta
_{q_{i_{1}}+...+q_{i_{s}}}R_{q_{i_{1}}...q_{i_{s}}}^{m_{i_{1}}...m_{i_{s}}}
\notag \\
&&\times \widetilde{\Gamma }%
_{k,q_{i_{s+1}}...q_{i_{n}}}^{m_{i_{s+1}}...m_{i_{n}},\sigma \sigma ^{\prime
}}  \notag
\end{eqnarray}%
where $I_{s}^{n}=\{i_{r}\}_{r=1}^{s}$ is the $s$-element ordered subset of $%
\{1...n\};$ $\{i_{r}\}_{r=s+1}^{n}=\{1...n\}\backslash I_{s}^{n},$ and the
summation is taken over all subsets $I_{s}^{n}$. The equations for the
vertices (\ref{conn}) read
\begin{eqnarray}
\widetilde{\Gamma }_{k;q_{1}...q_{n}}^{m_{1}...m_{n};\sigma \sigma ^{\prime
}} &=&TJ\sum_{m_{n+1}}{}\left\{ \widetilde{R}%
_{q_{1}...q_{n},q_{1}+...+q_{n}}^{m_{1}...m_{n+1}}\sigma _{\sigma \sigma
^{\prime }}^{m_{n+1}}\right.  \notag \\
&&-\sum_{s=1}^{n-1}\sum_{\{I_{s}\},\sigma ^{\prime \prime }}\widetilde{R}%
_{q_{i_{1}}...q_{i_{s}},q_{i_{1}}+...+q_{i_{s}}}^{m_{i_{1}}...m_{i_{s}},m_{n+1}}
\notag \\
&&\times G_{k+q_{i_{1}}+...+q_{i_{s}}}^{0}\widetilde{\Gamma }%
_{k,q_{i_{s+1}}...q_{i_{n}}}^{m_{i_{s+1}}...m_{i_{n}},\sigma \sigma ^{\prime
\prime }}\sigma _{\sigma ^{\prime \prime }\sigma ^{\prime }}^{m_{n+1}}
\notag \\
&&-\sum_{q_{n+1},\sigma ^{\prime \prime }}{}\widetilde{\Gamma }%
_{k;q_{1}...q_{n+1}}^{m_{1}...m_{n+1},\sigma \sigma ^{\prime \prime }}\sigma
_{\sigma ^{\prime \prime }\sigma ^{\prime }}^{m_{n+1}}  \notag \\
&&\ \ \times \left. G_{k+q_{1}+...+q_{n+1}}^{0}\right\}  \label{req}
\end{eqnarray}%
where $\widetilde{R}$ denotes the connected part of $R.$ The diagram
representation of the Eqs. (\ref{req}) for $n=1...3$ and the Eq. (\ref{SE})
for the self-energy is shown in Fig. \ref{Fig:Fig10}.

At the next step we express the vertices $\widetilde{\Gamma }%
_{k;q_{1}...q_{n}}^{m_{1}...m_{n}}$ through the corresponding one-particle
irreducible vertices $\gamma _{k;q_{1}...q_{n}}^{m_{1}...m_{n}}$ with the
use of the Legendre transformation of the generating functional, see, e.g.,
Ref. \cite{Ryder}. To the first order in $1/M$ the equations for $\Sigma
_{k},\gamma _{k,q_{1}},\gamma _{k;q_{1}q_{2}}^{zz},$ and $\gamma
_{k;q_{1}q_{2}q_{3}}^{zz\perp }$ form closed system, see Fig. 11. The
analytical form of these equations reads (below we omit the electronic spin
indices of $\gamma $ assuming that any of the equal nonzero spin components
of the vertex is taken):
\begin{eqnarray}
\Sigma _{k} &=&M\sum_{q}{}^{\prime }\gamma _{k;q}G_{k+q}\chi _{q}  \label{s1}
\\
\gamma _{k,q_{1}} &=&1+\sum_{q_{2}}{}^{\prime }\left[ (2-M)\gamma
_{k;q_{2}}\gamma _{k+q_{2};q_{1}}G_{k+q_{2}}\right.  \notag \\
&&\ \left. +\gamma _{k;q_{1},q_{2}}^{zz}\right] \chi
_{q_{2}}G_{k+q_{1}+q_{2}} \\
\gamma _{k;q_{1}q_{2}}^{zz} &=&M\sum_{q_{3}}{}^{\prime
}G_{k+q_{1}+q_{2}+q_{3}}\chi _{q_{3}}  \notag \\
&&\times \left[ \gamma _{k;q_{3}}\gamma _{k+q_{3};q_{1}}\gamma
_{k+q_{3}+q_{1};q_{2}}G_{k+q_{3}}G_{k+q_{3}+q_{1}}\right.  \notag \\
&&+\gamma _{k;q_{3}}\gamma _{k+q_{3};q_{2}}\gamma
_{k+q_{3}+q_{2};q_{1}}G_{k+q_{3}}G_{k+q_{3}+q_{2}}  \notag \\
&&+\gamma _{k,q_{1}q_{2}q_{3}}^{zz\perp }+\gamma _{k,q_{3}}\gamma
_{k+q_{3},q_{1}q_{2}}^{zz}G_{k+q_{3}}  \notag \\
&&\ \left. +M\gamma _{k,q_{1}+q_{2}+q_{3}}\Gamma
_{q_{1}q_{2}q_{3}}^{+-zz}\chi _{q_{1}+q_{2}+q_{3}}\right] \\
\gamma _{k;q_{1}q_{2}q_{3}}^{zz\perp } &=&M\sum_{q_{4}}{}^{\prime
}G_{k+q_{1}+q_{2}+q_{3}+q_{4}}\chi _{q_{4}}  \notag \\
&&\ \left[ \gamma _{k;q_{4}}\left( \gamma _{k+q_{4};q_{1}}\gamma
_{k+q_{4}+q_{1};q_{3}}\gamma _{k+q_{4}+q_{1}+q_{3};q_{2}}\right. \right.
\notag \\
&&\ \times G_{k+q_{4}}G_{k+q_{4}+q_{1}}G_{k+q_{4}+q_{1}+q_{3}}  \notag \\
&&\ +\gamma _{k+q_{4};q_{2}}\gamma _{k+q_{4}+q_{2};q_{3}}\gamma
_{k+q_{4}+q_{2}+q_{3};q_{1}}  \notag \\
&&\ \times G_{k+q_{4}}G_{k+q_{4}+q_{2}}G_{k+q_{4}+q_{2}+q_{3}}  \notag \\
&&\ -\gamma _{k+q_{4};q_{3}}\gamma _{k+q_{4}+q_{3};q_{1}}\gamma
_{k+q_{4}+q_{1}+q_{3};q_{2}}  \notag \\
&&\ \times G_{k+q_{4}}G_{k+q_{4}+q_{3}}G_{k+q_{4}+q_{3}+q_{1}}  \notag \\
&&\ -\gamma _{k+q_{4};q_{3}}\gamma _{k+q_{4}+q_{3};q_{2}}\gamma
_{k+q_{4}+q_{2}+q_{3};q_{1}}  \notag \\
&&\ \times G_{k+q_{4}}G_{k+q_{4}+q_{3}}G_{k+q_{4}+q_{3}+q_{1}}  \notag \\
&&\ -\gamma _{k+q_{4};q_{1}}\gamma _{k+q_{4}+q_{1};q_{2}}\gamma
_{k+q_{4}+q_{1}+q_{2};q_{3}}  \notag \\
&&\ \times G_{k+q_{4}}G_{k+q_{4}+q_{1}}G_{k+q_{4}+q_{1}+q_{2}}  \notag \\
&&\ -\gamma _{k+q_{4};q_{2}}\gamma _{k+q_{4}+q_{2};q_{1}}\gamma
_{k+q_{4}+q_{1}+q_{2};q_{3}}  \notag \\
&&\ \times G_{k+q_{4}}G_{k+q_{4}+q_{2}}G_{k+q_{4}+q_{1}+q_{2}}  \notag \\
&&\ -\gamma _{k+q_{4}+q_{1}+q_{2};q_{3}}\gamma
_{k+q_{4};q_{1}q_{2}}^{zz}G_{k+q_{4}}G_{k+q_{4}+q_{1}+q_{2}}  \notag \\
&&\ -\gamma _{k+q_{4};q_{3}}\gamma
_{k+q_{4}+q_{3};q_{1}q_{2}}^{zz}G_{k+q_{4}}G_{k+q_{4}+q_{3}}  \notag \\
&&\ \left. -\gamma _{k+q_{4};q_{1}q_{2}q_{3}}^{zz\perp }G_{k+q_{4}}\right)
-M\gamma _{k+q_{1}+q_{2}+q_{4};q_{3}}  \notag \\
&&\ \times \Gamma _{q_{1}q_{2}q_{4}}^{+-zz}G_{k+q_{1}+q_{2}+q_{4}}\chi
_{q_{1}+q_{2}+q_{4}}\gamma _{k,q_{1}+q_{2}+q_{4}}  \notag \\
&&\ \left. -\gamma _{k+q_{1}+q_{2}+q_{4};q_{3}}\gamma
_{k;q_{1}q_{2}q_{4}}^{zz\perp }G_{k+q_{1}+q_{2}+q_{4}}\right]  \label{g3}
\end{eqnarray}%
where $\chi _{q}=\chi _{q}^{z},$
\begin{eqnarray}
\Gamma _{q_{1}q_{2}q_{3}}^{m_{1}m_{2}m_{3}m_{4}} &=&TJ^{-2}\widetilde{R}%
_{q_{1}q_{2}q_{3},q_{1}+q_{2}+q_{3}}^{m_{1}m_{2}m_{3}m_{4}}  \notag \\
&&\times \,(\chi _{q_{1}}^{m_{1}}\chi _{q_{2}}^{m_{2}}\chi
_{q_{3}}^{m_{3}}\chi _{q_{1}+q_{2}+q_{3}}^{m_{4}})^{-1}  \label{Gamma4}
\end{eqnarray}%
is the irreducible vertex of the two-magnon interaction. In previous
treatments of the spin-fermion model (see, e.g., Ref. \cite{Schmalian}) the
vertex (\ref{Gamma4}) was neglected. For the flat density of states it
indeed vanishes in the limit of zero bosonic frequencies. Although it was
pointed out that this vertex is singular at nonzero external frequencies and
$T=0$ \cite{ChubukovSing}, this result is of possible relevance at finite
temperatures in the quantum-critical regime only, since in the renormalized
classical and high-temperature regimes the most important contribution to
the self-energy and vertices comes from the terms with zero bosonic
Matsubara frequencies. In this regimes the vertex (\ref{Gamma4}) may be
non-negligible for the non-constant (especially singular) density of states.
At the same time, the vertex (\ref{Gamma4}) does not vanish even for the
flat density of states in the $s$-$d$ model. To find this vertex we apply
the $1/M$-expansion \cite{CSY} to the spin part of the model (\ref{HH}). To
zeroth order in $1/M$ we find
\begin{equation}
\Gamma _{q_{1}q_{2}q_{3}}^{+-zz}=-2\left[ (JM)^{2}T\sum_{q_{^{\prime }}}\chi
_{q^{\prime }}\chi _{q^{\prime }+q_{1}+q_{2}}\right] ^{-1}
\end{equation}

\begin{figure}[tbp]
\includegraphics[width=8cm]{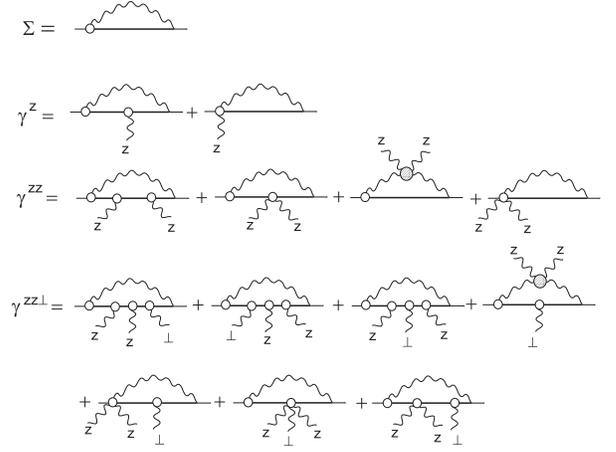} 
\caption{Diagram representation of the Eqs. (\protect\ref{s1})-(\protect\ref%
{Gamma4}). White circles stand for the one-particle irreducible vertices $%
\protect\gamma _{k;q_{1}...q_{n}}^{m_{1}...m_{n}}$, dashed circles for the
vertex $\Gamma _{q_{1}q_{2}q_{3}}^{+-zz}$, Eq. (\protect\ref{Gamma4}), bold
solid lines correspond to the full electronic Green function, wiggly lines -
to the magnetic susceptibility $\protect\chi _q$.}
\label{Fig:Fig11}
\end{figure}

To perform numerical calculations we consider two cases (see also the main
text): (i) the susceptibilities $\chi _{q}$ are static and
momentum-independent, $\chi _{q}=\delta _{n0}\chi ,$ the electronic Green's
functions also do not depend on the momentum, $G_{k}=G_{i\nu _{n}}$
(high-temperature regime); (ii) the susceptibilities have sharp maximum of
the width $\xi ^{-1}$ at $q=0$, the momentum and frequency dependence of the
electronic Green's functions being arbitrary (low-temperature regime).

In both the cases we approximate in the equation for the vertex $\gamma
_{k;q_{1}...q_{n}}^{m_{1}...m_{n}}$ ($n=0$ for the self-energy) the Green's
functions and vertices at $q_{i}\neq 0$ by the corresponding quantities at
the maximum total 4-momentum $k+q_{1}+...+q_{n},$ i.e. put
\begin{eqnarray}
\gamma _{k+\sum_{i}q_{i};Q_{1}...Q_{r}}^{m_{1}...m_{r}} &\simeq &\gamma
_{k+q_{1}+...+q_{n+1}}^{m_{1}...m_{r}}  \notag \\
G_{k+\sum_{i}q_{i}} &\simeq &G_{k+q_{1}+..+q_{n+1}}  \label{Approx1}
\end{eqnarray}%
In addition we use the approximation
\begin{eqnarray}
&&\sum_{q_{3}}f_{q_{i};q_{3}}\chi _{Q+q_{3}}\chi _{q_{3}}\Gamma
_{q_{1}q_{2}q_{3}}^{+-zz}%
\begin{array}{c}
\simeq%
\end{array}%
f_{q_{i};0}\sum_{q_{3}}\Gamma _{q_{1}q_{2}q_{3}}^{+-zz}\chi _{Q+q_{3}}\chi
_{q_{3}}  \notag \\
&&%
\begin{array}{c}
=%
\end{array}%
-2f_{q_{i};0}  \label{Approx2}
\end{eqnarray}%
where $f_{q_{i};q_{3}}$ is some function of electronic Green's functions and
vertices, and we have taken into account that $\Gamma
_{q_{1}q_{2}q_{3}}^{+-zz}$ depends in fact only on two first momenta.

Eqs. (\ref{Approx1}) and (\ref{Approx2}) are exact in the regime (i) where
the electronic Green's functions and vertices are momentum-independent and
the bosonic Matsubara frequencies can be put to zero. In the regime (ii) the
approximation (\ref{Approx1}) underestimates slightly the effect of magnetic
correlations, but treats them qualitatively correct and becomes exact in the
limit of infinite correlation length (where only vertices with the momentum $%
q=0$ enter). Approximation (\ref{Approx2}) is even more accurate in the
regime (ii), since the product $\chi _{Q+q_{3}}\chi _{q_{3}}$ is even
stronger peaked at $q_{3}=0.$ Applying (\ref{Approx1}) and (\ref{Approx2})
to (\ref{s1})-(\ref{g3}) we obtain the system of equations (\ref{Eqs}) of
the main text.

\end{document}